# Testing for Unobserved Heterogeneity via $k$-means Clustering[*]

Andrew J. Patton and Brian M. Weller

Duke University

15 July 2019

## Abstract

Clustering methods such as $k$-means have found widespread use in a variety of applications. This paper proposes a formal testing procedure to determine whether a null hypothesis of a single cluster, indicating homogeneity of the data, can be rejected in favor of multiple clusters. The test is simple to implement, valid under relatively mild conditions (including non-normality, and heterogeneity of the data in aspects beyond those in the clustering analysis), and applicable in a range of contexts (including clustering when the time series dimension is small, or clustering on parameters other than the mean). We verify that the test has good size control in finite samples, and we illustrate the test in applications to clustering vehicle manufacturers and U.S. mutual funds.

**J.E.L. codes:** C12, C38.

---

[*]For helpful comments we thank Tim Bollerslev and Jia Li. A supplemental appendix to this paper is available at http://www.econ.duke.edu/~ap172/research.html. Email: andrew.patton@duke.edu, brian.weller@duke.edu.

# 1 Introduction

Clustering methods provide researchers with a means of imposing some structure on a set of data under analysis. They represent a middle ground between imposing strict homogeneity and allowing complete heterogeneity across the variables under analysis, enabling the researcher to group variables into clusters and impose homogeneity only within a cluster. Such methods have proven useful in a wide variety of applications ranging including medical research (e.g., Eisen, *et al.*, 1998, and Liu, *et al.* 2008), economics (e.g., Francis, *et al.*, 2017, and Patton and Weller, 2019), and computer science (e.g., Ray and Turi, 1999, and Steinbach, *et al.*, 2000).

A key input to cluster analysis is the number of clusters to employ, and several methods for making this choice have been proposed in the literature. Perhaps most widely known is the "gap" statistic of Tibshirani *et al.* (2003), which looks at the reduction in a measure of within-cluster heterogeneity as a function of the number of clusters. Other approaches include those based on information criteria (e.g., Fraley and Raftery (2002), Sugar and James (2003) and Bonhomme and Manresa (2015)) and those based on cross-validation methods (e.g., Tibshirani and Walther (2005), Fu and Perry (2007) and Wang (2010)).

In many applications there is scientific interest in the null hypothesis of a *single* cluster, i.e., that the variables under analysis are homogeneous, or, more generally, homogeneous in the attribute(s) under analysis. A rejection of this hypothesis in favor of a model with multiple clusters represents evidence of heterogeneity, a conclusion that can have important implications. For example, a rejection could indicate that a medical treatment is effective only for some sub-populations; that investments with equal risk may have different expected returns; or that objects distinct from the background should trigger emergency application of the brakes. The methods for selecting the number of clusters described above do not allow for a probabilistic statement about the empirical evidence for or against a model with a single cluster. For that, we need a formal hypothesis test.

This paper proposes a general method for testing the null hypothesis of a single cluster imposing only mild regularity conditions on the data. We do so in the context of a panel of data containing $N$ variables, each with $T$ repeated observations, where the length of each dependent variable is $d$. Our testing approach exploits a standard assumption made in cluster analyses: cluster assignments are stable across repeated observations (e.g., time). This assumption enables us to estimate the cluster assignments on one sample (e.g., the first $T/2$ observations, or all odd-numbered observations) and then test the significance of the differences across clusters in a separate sample. This split-sample approach is simple to implement, and we show that it allows us to conduct inference under much weaker assumptions than existing methods. Our asymptotic theory is developed for $N, T \to \infty$, although we can also accommodate any fixed $T \geq 2$.

Some work has previously been done to test the significance of multiple clusters. Liu, *et al.* (2008) consider a high-dimensional setting $(d \gg N)$, and no repeated observations $(T = 1)$. Their approach takes a Gaussian distribution as the null hypothesis, which makes obtaining critical values for a test straightforward, however the assumption of Gaussianity is much stronger than the null of homogeneous means, and in many applications Gaussianity is not plausible. Maitra, *et al.* (2012) consider a bootstrap test for multiple clusters, replacing the assumption of Gaussianity with an assumption that the data are identically distributed after some known transformation. Our approach draws on recent work in panel econometrics to weaken these assumptions considerably: we impose *no* distributional assumptions on the data beyond standard regularity conditions and do not require homogeneity of the data beyond that implied by the clustering analysis.

The remainder of our paper is structured as follows. In Section 2 we present the main theoretical results, along with extensions to consider clustering on general estimated parameters (rather than means); tests when one of the clusters is "small;" and tests when the time series sample size is small. Section 3 presents simulation results on the finite-sample performance of the proposed methods, and Section 4 applies these tests to clustering vehicle manufacturers and U.S. mutual funds. Section 5 concludes. The appendix contains all proofs, and a web appendix contains additional details.

## 2 Testing for multiple clusters

Below we present our main result on testing for multiple clusters, followed by results related to the choice of $G$ under the alternative, and some empirically useful extensions of our main results.

## 2.1 Main result

We observe $T$ realizations of a collection of $N$ variables, $\mathbf{Y}_{it}$ for $i = 1, 2, ..., N$, and $t = 1, 2, ..., T$, where $\dim(\mathbf{Y}_{it}) = d$. In all cases we consider a split of the full sample of $T$ observations into two mutually exclusive, though not necessarily exhaustive, subsamples $\mathcal{R}$ and $\mathcal{P}$, where $\dim(\mathcal{R}) = R$ and $\dim(\mathcal{P}) = P$. Define $\mathcal{F}_R$ as the information set $\sigma\left(\{\mathbf{Y}_{it}\}_{i=1}^N, t \in \mathcal{R}\right)$.

**Assumption 1:** (a) The data come from $\mathbf{Y}_{it} = \mathbf{m}_i + \boldsymbol{\varepsilon}_{it}$, where $\boldsymbol{\varepsilon}_{it} = \Sigma_i^{1/2}\boldsymbol{\eta}_{it}$, $\boldsymbol{\eta}_{it} \sim iid\ F_i(\mathbf{0}, I_d)$, for $i = 1, ..., N$, and $t = 1, .., T$, where, for all $i$, $\mathbf{m}_i \in \mathcal{M} \subset \mathbb{R}^d$, $\Sigma_i$ is strictly positive definite, $E\left[\eta_{it}\eta_{jt}\eta_{kt}\eta_{lt}\right] \leq \bar{\kappa} < \infty\ \forall\ (i,j,k,l) \in \{1,..,N\}^4$, (b) $\boldsymbol{\eta}_{it} \perp\!\!\!\perp \boldsymbol{\eta}_{jt}\ \forall\ i \neq j$, (c) $N, P, R \to \infty$.

Assumption 1 allows the data to have arbitrary heterogeneity in variances and higher-order moments, subject to the existence of fourth-order moments. Importantly, it does not impose normality, as in Liu, *et al.* (2008), nor does it require the observations to be a known transformation away from homogeneity, as in Maitra, *et al.* (2012). Assumption 1 imposes that the data are independent across time and cross-sections; later in the paper we relax these conditions.

**Assumption 2:** $\mathbf{m}_i = \boldsymbol{\mu}^*\ \forall\ i$.

**Assumption 2':** For known $G \geq 2$, (a) $\mathbf{m}_i \in \{\boldsymbol{\mu}_1^*, ..., \boldsymbol{\mu}_G^*\}\ \forall\ i$, (b) $\left\|\boldsymbol{\mu}_g^* - \boldsymbol{\mu}_{g'}^*\right\| > c > 0\ \forall\ g \neq g'$, and (c) $\lim_{N\to\infty}\ N_g/N \equiv \pi_g \geq \underline{\pi} > 0$ for $g = 1, ..., G$, where $N_g \equiv \sum_{i=1}^N \mathbf{1}\{\gamma_i^* = g\}$, and $\gamma_i^* \in \{1, ..., G\}$ indicates to which cluster variable $i$ belongs.

Assumption 2 defines the homogeneous case we study under the null hypothesis. Assumption 2' covers the alternative hypothesis: (a) imposes that each variable belongs to one of the $G$ clusters, indicated by the group membership vector $\boldsymbol{\gamma}$, (b) imposes that the cluster means are "well separated," and (c) imposes that each cluster contains a non-trivial fraction of the total number of variables.

We stack the mean vectors for the $G$ clusters into a single $dG \times 1$ vector $\boldsymbol{\mu} \equiv [\boldsymbol{\mu}_1', ..., \boldsymbol{\mu}_G']'$. Define the full-sample estimator:

$$(\hat{\boldsymbol{\mu}}_{NT}, \hat{\boldsymbol{\gamma}}_{NT}) = \underset{(\boldsymbol{\mu}, \boldsymbol{\gamma}) \in \mathcal{M}^{dG} \times \Gamma_{N,G}}{\arg\min} \frac{1}{NT} \sum_{i=1}^N \sum_{t=1}^T \sum_{g=1}^G \left\|\mathbf{Y}_{it} - \boldsymbol{\mu}_g\right\|^2 \mathbf{1}\{\gamma_i = g\} \tag{1}$$

The set $\Gamma_{N,G}$ is the subset of all possible allocations of $N$ variables to $G$ groups that satisfies $\min_g \lim_{N\to\infty} N_g/N \geq \underline{\pi} > 0$, i.e., it only allows for "non-negligible" clusters.

Next define the estimator of the location parameters for a given value of $\boldsymbol{\gamma}$:

$$\tilde{\boldsymbol{\mu}}_{NT}(\boldsymbol{\gamma}) = \underset{\boldsymbol{\mu} \in \mathcal{M}^{dG}}{\arg\min} \frac{1}{NT} \sum_{i=1}^{N} \sum_{t=1}^{T} \sum_{g=1}^{G} \|\mathbf{Y}_{it} - \boldsymbol{\mu}_g\|^2 \mathbf{1}\{\gamma_i = g\} \tag{2}$$

We will look at a joint test that $\boldsymbol{\mu}_g^* = \boldsymbol{\mu}_{g'}^*$ for all $g \neq g'$, a total of $d(G-1)$ restrictions. To do so we will use the matrix:

$$\underset{(d(G-1)\times dG)}{A_{d,G}} = \left[(\iota_{G-1} \otimes I_d), -I_{d(G-1)}\right] \tag{3}$$

where $\iota_n$ is a $n \times 1$ vector of ones, $I_n$ is the $n \times n$ identity matrix, and $\otimes$ is the Kronecker product. This allows us to state the null as:

$$H_0 : A_{d,G}\boldsymbol{\mu}^* = 0 \Leftrightarrow H_0 : \boldsymbol{\mu}_g^* = \boldsymbol{\mu}_{g'}^* \ \forall \ g \neq g' \tag{4}$$

**Theorem 1** *Let $\hat{\boldsymbol{\gamma}}_{NR}$ be the estimated group assignments based on sample $\mathcal{R}$, and let $\tilde{\boldsymbol{\mu}}_{NP}(\hat{\boldsymbol{\gamma}}_{NR})$ be the estimated group means from sample $\mathcal{P}$ using group assignments $\hat{\boldsymbol{\gamma}}_{NR}$. Define the test statistic for the differences in the estimated means as*

$$F_{NPR} = NP\tilde{\boldsymbol{\mu}}'_{NP}(\hat{\boldsymbol{\gamma}}_{NR}) A'_{d,G} \left(A_{d,G}\hat{\Omega}_{NPR}A'_{d,G}\right)^{-1} A_{d,G}\tilde{\boldsymbol{\mu}}_{NP}(\hat{\boldsymbol{\gamma}}_{NR})$$

$$\text{where} \quad \underset{(dG\times dG)}{\hat{\Omega}^2_{NPR}} = \text{diag}\left\{\hat{\Omega}_{1,NPR}, ..., \hat{\Omega}_{G,NPR}\right\}$$

$$\text{and} \quad \underset{(d\times d)}{\hat{\Omega}_{g,NPR}} = \frac{1}{NP} \sum_{t\in\mathcal{P}} \sum_{i=1}^{N} (\mathbf{Y}_{it} - \bar{\mathbf{Y}}_{iP})(\mathbf{Y}_{it} - \bar{\mathbf{Y}}_{iP})' \hat{\pi}^{-2}_{g,NR} \mathbf{1}\{\hat{\gamma}_{i,NR} = g\}$$

$$\hat{\pi}_{g,NR} \equiv \frac{1}{N} \sum_{i=1}^{N} \mathbf{1}\{\hat{\gamma}_{i,NR} = g\}, \quad \text{for } g = 1, ..., G$$

*(a) Under Assumptions 1 and 2,*

$$F_{NPR} \xrightarrow{d} \chi^2_{d(G-1)}, \quad as \ N, P, R \to \infty$$

*(b) Under Assumptions 1 and 2',*

$$F_{NPR} \xrightarrow{p} \infty, \quad as \ N, P, R \to \infty$$

The proof is presented in the appendix. This theorem shows that if the means of the variables are homogeneous (i.e., Assumption 2 is satisfied) then the test statistic has a $\chi^2$ limit distribution, while if the variables are heterogeneous (Assumption 2′ is satisfied) then the test statistic diverges, and so this test has power to detect multiple groups.

Importantly, the null limiting distribution is not affected by the problem of estimated cluster assignments. Cluster assignments are unidentified under the null hypothesis, and obtaining results on the behavior of the estimated cluster assignments in such a case is difficult. Indeed, even when the clusters are well separated (i.e., under the alternative hypothesis), estimation error in cluster assignments is difficult to treat, see Pollard (1981, 1982) and Bonhomme and Manresa (2015). Without distribution theory for the estimated cluster assignments it is difficult to quantify the over-fitting problem that arises when estimating a multi-cluster model on homogeneous data, and simply ignoring the over-fitting problem leads to tests with poor size control: in the simulation study described in Section 3 we find rejection rates as high as 100% for a nominal 0.05 level test. Our test overcomes the overfitting problem via a simple split-sample approach.

Theorem 1 can be generalized to accommodate various departures from the assumptions given above. Time series dependence can be accommodated by employing results from Hansen (2007). The main change required when allowing for time series dependence is that the formation of subsamples ($\mathcal{R}$ and $\mathcal{P}$) now requires some structure. We suggest using simply the first and second halves of the time series. It is also possible to allow for general time series and cross-sectional dependence, drawing on results in Bonhomme and Manresa (2015) adapted to our application. The supplemental appendix contains details and formal results for these two extensions.

## 2.2 Choice of G under the alternative

The test above requires a choice of the number of clusters under the alternative, and in practice the value chosen may be incorrect. Below we consider the behavior of the test when the chosen value is too large or too small. The theory for behavior of the test statistic under the null is unaffected by this problem, of course, as under the null the true number of clusters is one and Theorem 1(a) applies. To simplify exposition, we assume that $d \equiv \dim(\mathbf{Y}_{it}) = 1$ in this section.

Firstly, consider the case that the model under the alternative ($\tilde{G}$) has more clusters than are needed ($G$). In this case the model considered under the alternative is "too big," but importantly it nests the correct model. We show below that the test remains consistent in this case, although in finite samples it may have lower power than the case where the correct value for the number of clusters is chosen. Consider an assumption based on the optimal $\tilde{G}$-cluster model:

**Assumption 3':** Assume $\tilde{G} > G > 1$, and (a) $p\lim_{N,R\to\infty} \hat{\boldsymbol{\mu}}_{NR}$ exists, and is denoted $\boldsymbol{\mu}^{\star}$. (b) $\min_g \lim_{N\to\infty} \tilde{N}_g/N \geq \underline{\pi} > 0$, where $\tilde{N}_g \equiv \sum_{i=1}^{N} \mathbf{1}\{\gamma_i^{\star} = g\}$, and $\gamma_i^{\star} \in \{1,...,\tilde{G}\}$ indicates to which cluster variable $i$ is assigned.

The lemma below shows that the optimal $\tilde{G}$-cluster parameter vector is the true $G$-cluster parameter vector, $\boldsymbol{\mu}^{*}$, with one or more of its elements repeated.

**Lemma 1** *Assume that the DGP satisfies Assumptions 1 and 2', but the researcher estimates a $\tilde{G} > G$ cluster model. Let $\boldsymbol{\mu}^{\star} = [\boldsymbol{\mu}^{*\prime}, \boldsymbol{\varphi}^{*\prime}]'$, where $\boldsymbol{\varphi}^{*}$ is a $(\tilde{G} - G)$ vector with elements drawn with replacement from $\boldsymbol{\mu}^{*}$, and let $\boldsymbol{\gamma}^{\star}$ be such that $\gamma_i^{\star} = \gamma_j^{\star} \Rightarrow \gamma_i^{*} = \gamma_j^{*} \ \forall \ i,j$. Then $(\boldsymbol{\mu}^{\star}, \boldsymbol{\gamma}^{\star})$ is a solution to the $\tilde{G}$-cluster model as $N,T \to \infty$.*

The proof is presented in the supplemental appendix. Lemma 1 reveals that under Assumption 2' the vector $\boldsymbol{\mu}^{\star}$ is "weakly well separated," in that $\left|\mu_g^{\star} - \mu_{g'}^{\star}\right| > c > 0$ for at least one pair $(g, g')$. In fact, this will hold for at least $(G-1)$ pairs $(g, g') \in \{1,...,\tilde{G}\}^2$. The presence of repeated values in $\boldsymbol{\mu}^{\star}$ means that some pair-wise differences will be zero.

Next consider the case that the model under the alternative ($\tilde{G}$) has fewer clusters than are needed ($G$). Choosing $\tilde{G}$ to be too small will generally mean that the estimated cluster means are not consistent for their true values, however our concern is only whether the null of a single cluster will be rejected. Assumption 3''(b) below states that the population values of the cluster means are, like the true cluster means, "well separated". Lemma 3 in the supplemental appendix shows that if $d = 1$ then $c^{\star} > c$, and so well-separatedness is ensured. For $d > 1$ it is possible to find cases where $c^{\star} < c$, and so in such cases we must simply assume the true cluster means are sufficiently well separated that the misspecified cluster means are also well separated.

**Assumption 3″:** Assume $G > \tilde{G} > 1$, and (a) $\mathrm{plim}_{N,R\to\infty} \hat{\boldsymbol{\mu}}_{NR}$ exists and is denoted $\boldsymbol{\mu}^\star$. (b) $\left|\mu_g^\star - \mu_{g'}^\star\right| > c^\star > 0 \;\forall\; g \neq g'$, and (c) $\min_g \lim_{N\to\infty} \tilde{N}_g/N \geq \underline{\pi} > 0$, where $\tilde{N}_g \equiv \sum_{i=1}^N \mathbf{1}\left\{\gamma_i^\star = g\right\}$, and $\gamma_i^\star \in \left\{1,..,\tilde{G}\right\}$ indicates to which cluster variable $i$ is assigned.

The following theorem contains results when the number of clusters under the alternative is larger or smaller than that chosen by the researcher.

**Theorem 2** *Let $\tilde{G}$ denote the number of groups considered by the researcher and let $\hat{\boldsymbol{\gamma}}_{NR}$ be the estimated group assignments based on sample $\mathcal{R}$, and let $\tilde{\boldsymbol{\mu}}_{NP}(\hat{\boldsymbol{\gamma}}_{NR})$ be the estimated group means from sample $\mathcal{P}$ using group assignments $\hat{\boldsymbol{\gamma}}_{NR}$. Define the test statistic for the differences in the estimated means as*

$$F_{NPR} = NP\tilde{\boldsymbol{\mu}}'_{NP}(\hat{\boldsymbol{\gamma}}_{NR}) A'_{\tilde{G}} \left(A_{\tilde{G}} \hat{\Omega}_{NPR} A'_{\tilde{G}}\right)^{-1} A_{\tilde{G}} \tilde{\boldsymbol{\mu}}_{NP}(\hat{\boldsymbol{\gamma}}_{NR})$$

*where* $\underset{(\tilde{G}\times\tilde{G})}{\hat{\Omega}_{NPR}} = \mathrm{diag}\left\{\hat{\omega}^2_{1,NPR},...,\hat{\omega}^2_{\tilde{G},NPR}\right\}$

*and* $\underset{(1\times 1)}{\hat{\omega}^2_{g,NPR}} = \dfrac{1}{NP} \sum_{t\in\mathcal{P}} \sum_{i=1}^N (Y_{it} - \bar{Y}_{iP})^2 \hat{\pi}^{-2}_{g,R} \mathbf{1}\{\hat{\gamma}_{i,NR} = g\}$

$$\hat{\pi}_{g,R} \equiv \frac{1}{N}\sum_{i=1}^N \mathbf{1}\{\hat{\gamma}_{i,NR} = g\}, \quad \text{for } g = 1,...,\tilde{G}$$

*(a) Under Assumptions 1 and 2,*

$$F_{NPR} \xrightarrow{d} \chi^2_{\tilde{G}-1}, \quad as\; N, P \to \infty$$

*(b) Under Assumptions 1, 2′ and 3′, or (c) 1, 2′ and 3″*

$$F_{NPR} \xrightarrow{p} \infty, \quad as\; N, P, R \to \infty$$

The proof is presented in the supplemental appendix. Theorem 2(b) shows that the test has unit asymptotic power under the alternative, even when $\tilde{G} > G$. In finite samples, power may be lower than if the correct number of clusters was used, as the critical values from a $\chi^2_{\tilde{G}}$ distribution are increasing in $\tilde{G}$. Theorem 2(c) confirms that if the cluster model with too few clusters is well separated, then we obtain the expected result for the test statistic under the alternative. We investigate the finite-sample impact of choosing an incorrect value of $\tilde{G}$ in Section 3.

With the results above we can consider a simple multiple testing procedure that applies when the researcher does not know the correct value for $G$ under the alternative, and wants to consider a range of possible values. For example, the researcher implements the test for $\tilde{G} = 2, ..., \bar{G}$, a total of $\bar{G}-1$ tests. The $p$-values from each of these tests, denoted $p_{\tilde{G}}$, can be combined via a Bonferroni adjustment: define the joint $p$-value as

$$p_{Bonf} = \min \left\{ 1, (\bar{G}-1) \times \min_{\tilde{G} \in \{2,...,\bar{G}\}} p_{\tilde{G}} \right\} \quad (5)$$

then reject the null that $G = 1$ in favor of $G \in \{2, ..., \bar{G}\}$ if $p_{Bonf} < \alpha$, where $\alpha$ is the desired level for the test. As usual with Bonferroni corrections, this procedure may be conservative under the null hypothesis. We investigate this in our simulation study in Section 3.

## 2.3 Extensions

### 2.3.1 Clustering on estimated parameters

Here we consider the problem of clustering on parameter, $\boldsymbol{\beta}_i \in \mathcal{A} \subset \mathbf{R}^b$, estimated for each of the $N$ variables. This allows researchers to cluster on features other than means, such as variances, other moments, regression coefficients, or other estimated parameters. We assume that the estimated parameter satisfies some standard regularity conditions, summarized in the following assumption.

**Assumption 4:**

(a) $\sqrt{T}\left(\hat{\boldsymbol{\beta}}_{i,T} - \boldsymbol{\beta}_i^*\right) \equiv \mathbf{Z}_{i,T}^* = \mathbf{Z}_{i,T} + \boldsymbol{\epsilon}_{i,T}$, where $\mathbf{Z}_{i,T} \sim N(\mathbf{0}, V_i)$ and $\boldsymbol{\epsilon}_{i,T} = o_p(1)$, for $i = 1, ..., N$.

(b) $\exists \hat{V}_{i,T}$ s.t. $\text{plim}_{T \to \infty} \hat{V}_{i,T} = V_i$, for $i = 1, ..., N$.

(c) $\mathbf{Z}_{i,T} \perp\!\!\!\perp \mathbf{Z}_{j,T} = 0 \ \forall \ i \neq j$ and $(\mathbf{Z}_{i,P}, \boldsymbol{\epsilon}_{i,P}) \perp\!\!\!\perp X \ \forall \ X \in \mathcal{F}_R \ \forall \ i$.

(d) $\frac{1}{N}\sum_{i=1}^{N} \boldsymbol{\epsilon}_{i,T} = o_p\left(N^{-1/2}\right)$

Assumption 4(a) requires that a standard first-order asymptotic Normal limit holds for the estimator, and 4(b) requires that a consistent estimator of the asymptotic variance is available. These assumptions are easily verified in a variety of different applications. Assumption 4(c) imposes that the first-order term in the estimation errors are uncorrelated in the cross-section, and imposes that estimation error from the $\mathcal{P}$ sample is independent of the $\mathcal{R}$ sample. The latter holds trivially if $\mathbf{Y}_{it}$ is $iid$ in the time series, but it also allows for some time series dependence, and the former can be relaxed to allow for mild cross-sectional correlation. Assumption 4(d) allows the higher-order terms in the estimation errors to have weak cross-sectional dependence.

The clustering model imposes

$$\boldsymbol{\beta}_i^* = \boldsymbol{\alpha}_{\gamma_i^0}^0 \quad \forall \ i = 1, ..., N \tag{6}$$

where $\boldsymbol{\alpha}_g^0$ is the cluster $g$ parameter. That is, the modeling assumption is that all variables in the same cluster have the same value for $\boldsymbol{\beta}_i^*$. We now modify Assumption 2 for this application:

**Assumption 2$_P$:** (a) The mean parameters satisfy $\boldsymbol{\beta}_i^* = \boldsymbol{\alpha}^* \ \forall \ i$.

**Assumption 2$'_P$:** (a) The mean parameters satisfy $\boldsymbol{\beta}_i^* \in \{\boldsymbol{\alpha}_1^*, ..., \boldsymbol{\alpha}_G^*\} \ \forall \ i$, (b) $\left|\boldsymbol{\alpha}_g^* - \boldsymbol{\alpha}_{g'}^*\right| > c > 0 \ \forall \ g \neq g'$, (c) $\lim_{N \to \infty} N_g/N \equiv \pi_g \geq \underline{\pi} > 0$ for $g = 1, ..., G$, where $N_g \equiv \sum_{i=1}^N \mathbf{1}\{\gamma_i^* = g\}$, and $\gamma_i^* \in \{1, ..., G\}$ indicates to which cluster variable $i$ belongs.

We stack the parameter vectors for the $G$ clusters into a single $bG \times 1$ vector $\boldsymbol{\alpha} \equiv [\boldsymbol{\alpha}_1', ..., \boldsymbol{\alpha}_G']'$ and define the full-sample estimators:

$$(\hat{\boldsymbol{\alpha}}_{NT}, \hat{\boldsymbol{\gamma}}_{NT}) = \underset{(\boldsymbol{\alpha}, \boldsymbol{\gamma}) \in \mathcal{A}^G \times \Gamma_{N,G}}{\arg\min} \frac{1}{N} \sum_{i=1}^N \sum_{g=1}^G \left(\hat{\boldsymbol{\beta}}_{i,T} - \boldsymbol{\alpha}_g\right)^2 \mathbf{1}\{\gamma_i = g\} \tag{7}$$

as well as the estimator of the cluster parameters for a given value of the group membership vector:

$$\tilde{\boldsymbol{\alpha}}_{NP}(\boldsymbol{\gamma}) = \underset{\boldsymbol{\alpha} \in \mathcal{A}^G}{\arg\min} \frac{1}{N} \sum_{i=1}^N \sum_{g=1}^G \left(\hat{\boldsymbol{\beta}}_{i,P} - \boldsymbol{\alpha}_g\right)^2 \mathbf{1}\{\gamma_i = g\} \tag{8}$$

The theorem provides a test for multiple clusters based on a general estimated parameter vector. The proof is presented in the appendix.

**Theorem 3** *Let $\hat{\boldsymbol{\gamma}}_{NR}$ be the estimated group assignments based on sample $\mathcal{R}$, and let $\tilde{\boldsymbol{\beta}}_{NP}(\hat{\boldsymbol{\gamma}}_{NR})$ be the estimated cluster parameters from sample $\mathcal{P}$ using group assignments $\hat{\boldsymbol{\gamma}}_{NR}$. Define the test* *statistic for the differences in the estimated means as*

$$F_{NPR} = NP\tilde{\alpha}'_{NP}(\hat{\gamma}_{NR}) A'_{b,G} \left(A_{b,G}\hat{\Omega}_{NPR}A'_{b,G}\right)^{-1} A_{b,G}\tilde{\alpha}_{NP}(\hat{\gamma}_{NR})$$

$$\text{where } \underset{(bG\times bG)}{\hat{\Omega}_{NPR}} = diag\left\{\hat{\Omega}_{1,NPR},...,\hat{\Omega}_{G,NPR}\right\}$$

$$\underset{(b\times b)}{\hat{\Omega}_{g,NPR}} = \frac{1}{N}\sum_{i=1}^{N} \hat{V}_{i,P}\hat{\pi}_{g,NR}^{-2}\mathbf{1}\left\{\hat{\gamma}_{i,NR} = g\right\}$$

$$\hat{\pi}_{g,NR} \equiv \frac{1}{N}\sum_{i=1}^{N} \mathbf{1}\left\{\hat{\gamma}_{i,NR} = g\right\}, \text{ for } g = 1,...,G$$

(a) Under Assumptions 4 and $2_P$,

$$F_{NPR} \xrightarrow{d} \chi^2_{b(G-1)}, \text{ as } N, P, R \to \infty$$

(b) Under Assumptions 4 and $2'_P$,

$$F_{NPR} \xrightarrow{p} \infty, \text{ as } N, P, R \to \infty$$

### 2.3.2 Dealing with "small" clusters

Our interest is in the joint restriction that $\mu_g^* = \mu_{g'}^*$ for all $g \neq g'$, a total of $(G-1)$ restrictions. To allow for the presence of "small" clusters, we will test an implication of this null, namely that $\mu_g^* = \mu_{g'}^*$ for all $g \neq g'$ s.t. $\pi_g, \pi_{g'} \geq \underline{\pi}$. We adjust Assumption 2(c) to require only that at least two clusters are "large." We simplify the exposition by assuming that $d \equiv \dim(Y_{it}) = 1$, but the results generalize naturally to the case that $d > 1$.

**Assumption 2'($\mathbf{c}^S$):** $\sum_{g=1}^{G} \mathbf{1}\{\pi_g \geq \underline{\pi}\} \geq 2$, where $\underline{\pi} > 0$, $\pi_g \equiv \lim_{N\to\infty} N_g/N \geq \underline{\pi} > 0$, $N_g \equiv \sum_{i=1}^{N} \mathbf{1}\{\gamma_i^* = g\}$, and $\gamma_i^* \in \{1,..,G\}$ indicates to which cluster variable $i$ belongs.

To implement this test, order the clusters so that $\hat{\pi}_{1,NR} \geq \hat{\pi}_{2,NR} \geq \cdots \geq \hat{\pi}_{G,NR}$, and define

$$\hat{G}_{NR} = \max_g \hat{\pi}_{g,NR} \text{ s.t. } \hat{\pi}_{g,NR} \geq \underline{\pi} \tag{9}$$

That is, $\hat{G}_{NR}$ is the estimated number of "large" clusters. For $2 \leq G' \leq G$, define the matrix

$$\underset{((G'-1)\times G)}{B_{G',G}} = \left[\iota_{G'-1}, -I_{(G'-1)}, \mathbf{0}_{(G'-1,G-G')}\right] \tag{10}$$

This is the matrix comprised of the first $(G'-1)$ rows of $A_{1,G}$ defined in equation (3) above. This allows us to obtain an implication of the null for the $\hat{G}_{NR}$ "large" clusters:

$$H_0^S : B_{\hat{G}_{NR},G}\boldsymbol{\mu}^* = 0 \tag{11}$$

Note that below we characterize the asymptotic distribution of the $p$-value of the test statistic rather than the test statistic itself. The limiting distribution of the latter depends on the value for $\hat{G}_{NR}$, which in turn depends on $\mathcal{F}_R \equiv \sigma\left(\{\mathbf{Y}_{it}\}_{i=1}^N, t \in \mathcal{R}\right)$. Our proof technique relies on the limiting distribution being independent of $\mathcal{F}_R$; we achieve this below by transforming the test statistic to a $p$-value.

**Theorem 4** *Let $\hat{\boldsymbol{\gamma}}_{NR}$ be the estimated group assignments based on sample $\mathcal{R}$, and let $\tilde{\boldsymbol{\mu}}_{NP}(\hat{\boldsymbol{\gamma}}_{NR})$ be the estimated group means from sample $\mathcal{P}$ using group assignments $\hat{\boldsymbol{\gamma}}_{NR}$. Let $\Upsilon(\cdot;q)$ denote the CDF of a $\chi^2$ variable with $q$ degrees of freedom, and define the p-value for the differences in the estimated means as:*

$$\begin{aligned}
Pval_{NPR} &= 1 - \Upsilon\left(F_{NPR}; \hat{G}_{NR}-1\right) \\
\text{where } F_{NPR} &= NP\tilde{\boldsymbol{\mu}}'_{NP}(\hat{\boldsymbol{\gamma}}_{NR}) B'_{\hat{G}_{NR},G} \left(B_{\hat{G}_{NR},G}\hat{\Omega}_{NPR}B'_{\hat{G}_{NR},G}\right)^{-1} B_{\hat{G}_{NR},G}\tilde{\boldsymbol{\mu}}_{NP}(\hat{\boldsymbol{\gamma}}_{NR}) \\
\underset{(dG \times dG)}{\hat{\Omega}_{NPR}^2} &= diag\left\{\hat{\Omega}_{1,NPR}, ..., \hat{\Omega}_{G,NPR}\right\} \\
\underset{(d \times d)}{\hat{\Omega}_{g,NPR}} &= \frac{1}{NP}\sum_{t \in \mathcal{P}}\sum_{i=1}^{N}\left(\mathbf{Y}_{it}-\bar{\mathbf{Y}}_{iP}\right)\left(\mathbf{Y}_{it}-\bar{\mathbf{Y}}_{iP}\right)' \hat{\pi}_{g,R}^{-2}\mathbf{1}\left\{\hat{\gamma}_{i,NR}=g\right\} \\
\hat{\pi}_{g,R} &\equiv \frac{1}{N}\sum_{i=1}^{N}\mathbf{1}\left\{\hat{\gamma}_{i,NR}=g\right\}, \text{ for } g=1,...,G
\end{aligned}$$

*(a) Under Assumptions 1 and 2,*

$$Pval_{NPR} \xrightarrow{d} Unif(0,1), \text{ as } N,P,R \to \infty$$

*(b) Under Assumptions 1 and $2'(a),(b),(c^S)$,*

$$Pval_{NPR} \xrightarrow{p} 0, \text{ as } N,P,R \to \infty$$

### 2.3.3 Diverging N and finite T

We consider here the case that the number of repeated observations ($T$, in our notation) is small relative to the number of variables, $N$. Our split-sample approach to overcome the over-fitting problem requires only $T \geq 2$, not $T \to \infty$. We consider the finite $T$ case by modifying Assumption 1 as follows. We again simplify exposition by assuming that $d \equiv \dim(\mathbf{Y}_{it}) = 1$ here, but the results generalize naturally to the case that $d > 1$.

**Assumption 1':** (a) The data come from $Y_{it} = m_i + \varepsilon_{it}$, for $i = 1, ..., N$, and $t = 1, .., T \geq 2$, where $m_i \in [\underline{m}, \bar{m}] \subset \mathbb{R}$ and $V[\varepsilon_{it}] \equiv \sigma_i^2 \in [\underline{\sigma}^2, \bar{\sigma}^2] \subset \mathbb{R}_{++} \ \forall\ i$, $E[\varepsilon_{it}] = 0$ and $E\left[|\varepsilon_{it}|^{4+\delta}\right] < \infty \ \forall\ i$ for some $\delta > 0$, (b) $\varepsilon_{it} \perp\!\!\!\perp \varepsilon_{jt} \ \forall\ i \neq j$, and $\varepsilon_{it} \perp\!\!\!\perp \varepsilon_{js} \ \forall\ i, j$ for $(t, s) \in \{\mathcal{R}, \mathcal{P}\}$, and (c) $N \to \infty$, and $R, P \geq 1$.

Assumption 1'(a) allows for cross-sectional heteroskedasticity, and heterogeneity more generally, in the distribution of residuals, subject to them being mean zero and having finite $4 + \delta$ moments. Assumption 1'(b) imposes cross-sectional independence, and time series independence across the $\mathcal{R}$ and $\mathcal{P}$ subsamples. Within each of the $\mathcal{R}$ and $\mathcal{P}$ subsamples, time series dependence is not constrained. Assumption 1'(c) requires the cross-sectional dimension to diverge, and each subsample to have at least one observation.

**Theorem 5** *Let $\hat{\boldsymbol{\gamma}}_{NR}$ be the estimated group assignments based on sample $\mathcal{R}$, and let $\tilde{\mu}_{NP}(\hat{\boldsymbol{\gamma}}_{NR})$ be the estimated group means from sample $\mathcal{P}$ using group assignments $\hat{\boldsymbol{\gamma}}_{NR}$. Define the t-statistic for the differences in the estimated means as:*

$$tstat_{NPR} = \frac{\sqrt{NP}\left(\tilde{\mu}_{1,NP}(\hat{\boldsymbol{\gamma}}_{NR}) - \tilde{\mu}_{2,NP}(\hat{\boldsymbol{\gamma}}_{NR})\right)}{\hat{\omega}_{NPR}}$$

$$\text{where}\quad \hat{\omega}_{NPR}^2 \equiv \frac{1}{NP}\sum_{i=1}^{N} \boldsymbol{\iota}_P' \hat{\boldsymbol{\varepsilon}}_i \hat{\boldsymbol{\varepsilon}}_i' \boldsymbol{\iota}_P \left(\hat{\pi}_{1,NR}^{-2} \mathbf{1}\{\hat{\gamma}_{i,NR} = 1\} + \hat{\pi}_{2,NR}^{-2} \mathbf{1}\{\hat{\gamma}_{i,NR} = 2\}\right)$$

$$\underbrace{\hat{\boldsymbol{\varepsilon}}_i}_{(P\times 1)} = \underbrace{\mathbf{Y}_i}_{(P\times 1)} - \underbrace{\tilde{\mu}_{\hat{\gamma}_{i,NR},NP}(\hat{\boldsymbol{\gamma}}_{NR})}_{(1\times 1)}$$

$$\text{and}\quad \hat{\pi}_{g,NR} \equiv \frac{1}{N}\sum_{i=1}^{N} \mathbf{1}\{\hat{\gamma}_{i,NR} = g\},\quad \text{for } g = 1, 2$$

(a) Under Assumptions 1' and 2,

$$tstat_{NPR} \xrightarrow{d} N(0,1), \quad as\ N \to \infty$$

(b) Under Assumptions 1' and 2',

$$|tstat_{NPR}| \xrightarrow{p} \infty, \quad as\ N \to \infty$$

This theorem expands the applicability of the testing approach proposed in this paper: we now only need $T \geq 2$, rather than $T$ "large" in an asymptotic sense. Of course, the power of the test will be greater if a larger sample size is available, but this theorem shows that even in applications with a small time series sample size, the proposed testing approach may be adopted.

## 3 Simulation study

In this section we investigate the finite-sample behavior of the proposed tests. We first study the finite-sample size of the test, using the design:

$$\mathbf{Y}_{it} = \mathbf{m}_i + \boldsymbol{\varepsilon}_{it}, \ i = 1, ..., N; \ t = 1, ..., T \tag{12}$$

$$\boldsymbol{\varepsilon}_{it} \sim iid\ F_i(\mathbf{0}, I_d)$$

We impose $\mathbf{m}_i = \mathbf{0}\ \forall\ i$, thereby ensuring that the null of homogeneous means is satisfied. We consider a variety of configurations of the problem: $N \in \{30, 150, 300\}$, $T \in \{50, 250, 1000\}$, $d \in \{1, 2, 5\}$, $G \in \{2, 3, 4, 5\}$. In addition to the four individual values of $G$ considered under the alternative, we also study the performance of a Bonferroni-corrected combination method that considers all four tests.

We take $\boldsymbol{\varepsilon}_{it}$ to be Normally distributed or heterogeneously distributed; in the latter case the distribution for each variable $i$ is randomly selected from one of $N(0,1)$, $Exp(2)$, $Unif(-3,3)$, $\chi^2(4)$ or $t(5)$, standardized to have zero mean and unit variance. The heterogeneous data cannot be considered using the tests of Liu, *et al.* (2008) and Maitra, *et al.* (2012). We implement the test in Theorem 1 at the 0.05 significance level, splitting the time series evenly to form the $\mathcal{R}$ and $\mathcal{P}$ samples. We use 1000 replications.

Table 1 reports the finite-sample size results. We see that the rejection rates are generally very close to the nominal level of 0.05, for both the Normal and the heterogeneous data. In the supplemental appendix we repeat this simulation study using a test that does *not* split the time series into $\mathcal{R}$ and $\mathcal{P}$ samples. Table SA.1 reveals that the finite-sample rejection rates for such an approach are 100% in all but one configuration (where it is instead 99%), confirming the finite-sample size problems stemming from $k$-means overfitting the data, and motivating an approach such as ours.

[ INSERT TABLE 1 ABOUT HERE ]

We next consider the finite-sample power of the proposed test. We fix $d \equiv \dim(\mathbf{Y}_{it}) = 1$ and we consider an alternative containing $G = 2$ clusters. The cluster means are set to $(0, \mu_2)$, with $\mu_2 \in [0, 0.5]$. The case that $\mu_2 = 0$ corresponds to the null of a single cluster, and the rejection rate at that point should equal 0.05, the size of the test. As $\mu_2$ increases the cluster means become better separated and we expect the test to reject the null with greater frequency. Figure 1 reveals that the test has strong power to reject the null hypothesis when the sample sizes $(N, T)$ are large, and when the distance between cluster means is large. When $(N, T) = (30, 50)$ the test fails to detect small differences between the cluster means, and unit power is only achieved when $\mu_2 = 0.5$. When $(N, T) = (150, 1000)$ even small differences are significant, and unit power is achieved at $\mu_2 = 0.1$. It is noteworthy that the power of the test is essentially identical for Normally and heterogeneously distributed data. For the remainder of the simulation results we focus on Normally distributed data; the results for heterogeneously distributed data are very similar.

[ INSERT FIGURE 1 ABOUT HERE ]

Figure 2 studies the sensitivity of the test to the choice of number of clusters under the alternative. We consider two representative combinations of sample sizes $(N, T) = (30, 50)$ and $(150, 250)$, and for each sample size pair we choose a value of $\mu_2$ such that the test has power strictly inside $(0.05, 1)$, namely $\mu_2 = 0.2$ and $\mu_2 = 0.075$ respectively. In the left panel, the true number of clusters is two, and we consider tests that allow for between two and five clusters under the alternative.

Consistent with intuition, for both sample size pairs, we observe a decrease in power as the number of clusters is increased from two to five, though the decrease is small (e.g., power drops from 0.21 to 0.20 for the smaller sample size). In the right panel the true number of clusters is five (with cluster means evenly spaced between zero and either 0.2 or 0.075 depending on the sample size). Like the left panel, we find that power is nearly unaffected by the choice of $G$, with a slight increase in power from using smaller $G$. Though the models with $G < 5$ are misspecified, reducing the fit and thus the power, Lemma 3 shows that the too-small models will have cluster means that are better separated than the correct model, increasing power. Overall, Figure 2 suggests that the test exhibits robustness to the choice of $G$.

Figure 3 examines the performance of a test based on a Bonferroni adjustment to combine four tests using $G = 2, 3, 4, 5$, compared with a test that correctly chooses $G = 2$. Unsurprisingly, the Bonferroni-corrected test is conservative, and exhibits lower power than the test using the correct value of the $G$. When the sample sizes are small, $(N,T) = (30, 50)$, the lower is power is sizeable, however for larger sample sizes, e.g. $(N,T) = (150, 250)$, the power loss is minimal.

[ INSERT FIGURES 2 AND 3 ABOUT HERE ]

Next we study the performance of the test in Theorem 4, designed to accommodate small clusters. We again consider $(N,T) = (30, 50)$ and $(150, 250)$, with $d = 1$. We set the number of clusters to three, and we look at the impact of a small cluster by varying the proportion of variables in the third cluster, denoted $\pi_3$. We set $\pi_1 = \pi_2 = (1 - \pi_3)/2$, and consider $\pi_3 \in [1/100, 1/3]$, with the largest value for $\pi_3$ corresponding to all clusters having the same weight. We set the mean of the first cluster to zero in all cases, $\mu_1 = 0$, and we set the second cluster mean $\mu_2 = \mu_3/2$. To study the finite-sample size of the test, we set $\mu_3 = 0$. To study power we choose $\mu_3$ such that the test has power strictly inside $(0.05, 1)$, namely $\mu_3 = 0.2$ for $(N,T) = (30, 50)$ and $\mu_3 = 0.075$ for $(N,T) = (150, 250)$. We use a threshold of $\underline{\pi} = 0.1$ to decide whether a cluster is "small" and thus excluded from the test. The left panel of Figure 4 shows that the test in Theorem 4 controls the size of the test. The right panel shows, as expected, that the power of the test increases as the smallest cluster grows to be closer in size to the other two clusters.

To illustrate the applicability of the test for clusters on a general estimated parameter, we next consider an application where the clusters are found using the variables' autoregressive coefficients. That is, for each variable $Y_{i,t}$ we consider the autoregression:

$$Y_{i,t} = \phi_{0,i} + \phi_{1,i} Y_{i,t-1} + \varepsilon_{i,t} \tag{13}$$

and the cluster model assumption is that the AR(1) coefficients take one of only two values

$$\phi_{1,i} = \alpha_{\gamma_i}, \ \ i = 1, 2 \tag{14}$$

We fix $\alpha_1 = 0.5$ and we vary the autoregressive coefficient of the second cluster, $\alpha_2 \in [0.1, 0.9]$. Figure 5 shows the results for two representative combinations of sample sizes, $(N, T) = (30, 50)$ and $(150, 250)$. For the smaller sample size the test is unable to reject the null of a single cluster for values of $\alpha_2$ within about 0.1 of $\alpha_1$; the sampling variation in the estimated AR(1) parameters is simply too large in that case. As the differences between the cluster AR(1) parameters grows, or if we use a larger sample size, the power of the test increases. For both sample sizes the finite-sample size of the test is close to the nominal value.

[ INSERT FIGURES 4 AND 5 ABOUT HERE ]

Finally, we investigate the performance of the test in Theorem 5, which is applicable when $T$ is small. We consider $T \in \{2, 4, 6, 10\}$, and values of $N \in \{30, 150, 600\}$. Figure 6 shows that even when $T = 2$, the test has reasonable size control: the rejection rate for $N = 30$ is 0.07, so only slightly above the nominal level. (The rejection rates when $N = 30$ and $T = 4$, 6, or 10 are between 0.07 and 0.08.) The power is low for the smallest value of $N$, but when $N = 150$ or 600 power is non-trivial. As $T$ increases to 4, 6 and 10 we see that size control is maintained, and power increases. Naturally, a test with such small values of $T$ has lower power than for larger values of $T$, e.g. the results in Figure 1, however the results in Figure 6 show that even for very small values of $T$, size control is maintained and non-trivial power can be achieved with a large cross-sectional sample size.

[ INSERT FIGURE 6 ABOUT HERE]

# 4 Empirical applications

## 4.1 Vehicle manufacturer clusters

To illustrate our methodology in a well-known setting, we use a standard data set, built into MATLAB, on car attributes for 307 vehicle models from 30 manufacturers in seven countries during 1970–1982. Vehicle attributes include acceleration, number of cylinders, engine displacement, horsepower, miles per gallon, and weight, the last of which we log to avoid identifying outliers as separate clusters. We split our data into $\mathcal{R}$ and $\mathcal{P}$ samples of 1970–1975 and 1976–1982, respectively. Within each sample, we average vehicle attributes by manufacturer across all combinations of models and model years, and we retain only observations with non-missing values of all attributes and manufacturers with models in both samples. Our resulting sample consists of 24 manufacturers. To make scales comparable across characteristics, we standardize each attribute within each sample by demeaning and dividing by the standard deviation across manufacturers.

We assuming $G = 2$ clusters, and use $k$-means on the $\mathcal{R}$ sample, with $1,000$ starting values initialized by $k$-means++ (Arthur and Vassilvitskii, 2007). Table 2 summarizes the results. Panel A shows that "Group 1" manufacturers typically produce vehicles with more cylinders, larger engines, greater horsepower, lower mileage, and greater weight than "Group 2" manufacturers. Note that all cross-cluster characteristic differences are larger on the $\mathcal{R}$ sample than on the $\mathcal{P}$ sample, consistent with the clustering procedure fitting both true differences among manufacturers as well as noise. The $p$-value from the test for multiple clusters is less than 0.001, indicating strong evidence in against the null of a single cluster. We conclude that at least two clusters are needed to describe vehicle manufacturers during this period.

In Panel B of Table 2 we present the constituents of each cluster, and a clear pattern emerges: we find that manufacturers cleave perfectly by region of origin, with Group 1 comprised completely of American manufacturers, and Group 2 containing all non-American manufacturers. While this dimension of heterogeneity may have been conjectured *ex ante*, the new test reveals that this heterogeneity is significant even controlling for all other possible splits that could be considered.

[ INSERT TABLE 2 ABOUT HERE ]

## 4.2 Mutual fund clusters

Performance evaluation, e.g. for mutual funds or hedge funds, is one of the central concerns of empirical finance. Most performance evaluation takes the form of comparing fund returns to a benchmark return, e.g., the return on a strategy or style with similar risk characteristics. A popular paper in style analysis, Brown and Goetzmann (1997) pioneered the application of $k$-means clustering for the purpose of benchmark formation and assignment of funds to benchmarks. We use the testing approach proposed in this paper to determine whether mutual fund styles are truly distinct in the data. We cluster based on risk exposures (betas) rather than returns themselves (as done in the original study) to facilitate interpretation of the results.

We use daily data from the CRSP Mutual Fund Database, see Patton and Weller (2019) for the data construction, filtering, and aggregation methodology. We use the first full year of the daily series (1999) for the $\mathcal{R}$ sample and the second year (2000) for the $\mathcal{P}$ sample, and we retain only U.S. domestic equity mutual funds that report for at least half the days in each year. The resulting sample consists of 1,743 mutual funds.

We run the following regression for each fund:

$$r_{it} = \alpha_i + \sum_{k=1}^{4} \beta_{ik} f_{kt} + \sigma_i \varepsilon_{it} \quad (15)$$

As factors, $f_{kt}$, we use the value-weighted market ($MKT$), size ($SMB$), value ($HML$), and momentum ($UMD$) returns of the Carhart (1997) model.[1] We also estimate average abnormal returns ($\alpha_i$) and idiosyncratic volatility ($\sigma_i$) for each fund but we do not cluster on these attributes.

We use $k$-means clustering on the $\mathcal{R}$ sample, with 1,000 starting values initialized by $k$-means++. We follow Brown and Goetzmann (1997) and use $G = 8$ clusters. Table 3 summarizes the results of the clustering procedure. Fund clusters differ markedly in the parameters on which the clustering was done (the risk exposures, $\beta_{ik}$) and interestingly also in the other parameters of the model ($\alpha_i$ and $\sigma_i$). For example, annualized average abnormal returns ($\alpha_i$) range between -3% and 22% across the clusters. This heterogeneity cannot be accommodated by other tests for multiple clusters.

---

[1] The factor data is available at http://mba.tuck.dartmouth.edu/pages/faculty/ken.french/data_library.html.

Unlike the two-group example in the previous section, differences between these eight groups, each with four dimensions of characteristics, are more difficult to present in tabular form. Nevertheless, the factor loadings in Table 3 reveal some clear clusters: Group 1, with a loading of near one of the market factor and relatively small loadings on the other three factors, is a "market" style cluster; Group 2, with high loadings on both the market and the size factor, is a "small capitalization" style cluster; Group 7, with high loadings on the aggregate market and value factors, is a value cluster; and Group 8, with factor loadings close to zero on all four factors, is a "market-neutral" style cluster. The $p$-value from the test for multiple clusters is less than 0.001, indicating strong evidence against the null of a single cluster. We conclude that mutual funds indeed have different styles.

[ INSERT TABLE 3 ABOUT HERE ]

# 5 Conclusion

This paper proposes methods to determine whether a null hypothesis of a single cluster, indicating homogeneity of the data, can be rejected in favor of multiple clusters. The new test is simple to implement and valid under relatively mild conditions, including non-normality, and heterogeneity of the data in aspects beyond those in the clustering analysis. We show via an extensive simulation study that the test has good finite-sample size control. We present extensions of the test for a range of applications, including clustering when the time series dimension is small, or clustering on parameters other than the mean. Some interesting extensions remain. For example, García-Escudero and Gordaliza (1999) propose a robust version of $k$-means based on trimmed means, Witten and Tibshirani (2010) propose a method to optimally choose the features on which to cluster when a large number of features is available, and Ng, *et al.* (2002) propose a spectral clustering method. We leave the analysis of these interesting extensions for future research.

# Appendix: Proofs

**Proof of Theorem 1.** (a) We first find the limiting distribution of $\sqrt{NP}\tilde{\boldsymbol{\mu}}_{NP}(\hat{\gamma}_{NR})$ conditional on $\mathcal{F}_R$. Denote $\hat{N}_{g,R} \equiv \frac{1}{N}\sum_{i=1}^{N} \mathbf{1}\{\hat{\gamma}_{g,NR} = 1\}$ and $\hat{\pi}_{g,R} = \hat{N}_{g,R}/N$, and note that

$$\tilde{\boldsymbol{\mu}}_{g,NP}(\hat{\gamma}_{NR}) = \frac{1}{\hat{N}_{g,R}}\sum_{i=1}^{N}\left(\mathbf{1}\{\hat{\gamma}_{i,NR} = g\}\frac{1}{P}\sum_{t\in\mathcal{P}}\mathbf{Y}_{i,t}\right)$$

$$= \frac{1}{NP}\sum_{i=1}^{N}\sum_{t\in\mathcal{P}}\mathbf{Y}_{i,t}\hat{\pi}_{g,R}^{-1}\mathbf{1}\{\hat{\gamma}_{i,NR} = g\}$$

for $g = 1, ..., G$. Thus

$$\sqrt{NP}\left(\tilde{\boldsymbol{\mu}}_{g,NP}(\hat{\gamma}_{NR}) - \boldsymbol{\mu}_g^*\right) = \frac{1}{\sqrt{NP}}\sum_{i=1}^{N}\sum_{t\in\mathcal{P}}(\boldsymbol{\mu}_g^* + \boldsymbol{\varepsilon}_{it})\hat{\pi}_{g,R}^{-1}\mathbf{1}\{\hat{\gamma}_{i,NR} = g\} - \sqrt{NP}\boldsymbol{\mu}_g^*$$

$$= \frac{1}{\sqrt{NP}}\sum_{i=1}^{N}\sum_{t\in\mathcal{P}}\hat{U}_{ig,NR}\boldsymbol{\varepsilon}_{it}$$

where $\hat{U}_{ig,NR} \equiv \hat{\pi}_{g,R}^{-1}\mathbf{1}\{\hat{\gamma}_{i,NR} = g\}$. Note that this variable is bounded as $\hat{\pi}_{g,R} \geq \underline{\pi} > 0$. Conditional on $\mathcal{F}_R$, the sequence $\{\hat{U}_{ig,NR}\boldsymbol{\varepsilon}_{it}\}$ is independent and heterogeneously distributed. Define

$$\bar{\Omega}_{gNR} = V\left[\frac{1}{\sqrt{NP}}\sum_{i=1}^{N}\sum_{t\in\mathcal{P}}\hat{U}_{ig,NR}\boldsymbol{\varepsilon}_{it}\middle|\mathcal{F}_R\right] = \frac{1}{N}\sum_{i=1}^{N}\hat{U}_{ig,NR}^2\Sigma_i$$

where $\Sigma_i \equiv V[\boldsymbol{\varepsilon}_{it}]$ and the second equality holds as $\boldsymbol{\varepsilon}_{it}$ is uncorrelated in the time series and cross section. Combining the Cramér-Wold device with a central limit theorem for *inid* random variables (e.g., Theorem 5.11 of White, 2001), we obtain the asymptotic distribution of $\tilde{\boldsymbol{\mu}}_{g,NP}(\hat{\gamma}_{NR})$:

$$\sqrt{NP}\bar{\Omega}_{gNR}^{-1/2}\left(\tilde{\boldsymbol{\mu}}_{g,NP}(\hat{\gamma}_{NR}) - \boldsymbol{\mu}_g^*\right) \xrightarrow{d} N(0, I)$$

This holds for each $g = 1, .., G$. Next we show that $Cov\left[\tilde{\boldsymbol{\mu}}_{g,NP}(\hat{\gamma}_{NR}), \tilde{\boldsymbol{\mu}}_{g',NP}(\hat{\gamma}_{NR})\right] = 0$ for all $g \neq g'$. Define $\bar{\varepsilon}_{ikP} \equiv \frac{1}{P}\sum_{t\in\mathcal{P}}\varepsilon_{itk}$, and consider elements $(k, k')$ of the vector $\left(\tilde{\boldsymbol{\mu}}_{g,NP}(\hat{\gamma}_{NR}) - \boldsymbol{\mu}_g^*\right)$. The covariance between any two elements $(k, k')$ in groups $g \neq g'$ is

$$E\left[\left(\tilde{\mu}_{gk,NP}(\hat{\gamma}_{NR}) - \mu_{gk}^*\right)\left(\tilde{\mu}_{g'k',NP}(\hat{\gamma}_{NR}) - \mu_{g'k'}^*\right)\middle|\mathcal{F}_R\right]$$

$$= \frac{1}{N^2}E\left[\left(\sum_{i=1}^{N}\hat{\pi}_{g,R}^{-1}\mathbf{1}\{\hat{\gamma}_{i,NR} = g\}\bar{\varepsilon}_{ikP}\right)\left(\sum_{j=1}^{N}\hat{\pi}_{g',R}^{-1}\mathbf{1}\{\hat{\gamma}_{j,NR} = g'\}\bar{\varepsilon}_{jk'P}\right)\middle|\mathcal{F}_R\right]$$

$$= 0$$

since $\mathbf{1}\{\hat{\gamma}_{i,NR} = g\}\mathbf{1}\{\hat{\gamma}_{i,NR} = g'\} = 0$ for $g \neq g'$. Thus we obtain the limiting distribution for the entire vector $\tilde{\boldsymbol{\mu}}_{NP}(\hat{\gamma}_{NR})$:

$$\sqrt{NP}\bar{\Omega}_{NR}^{-1/2}(\tilde{\boldsymbol{\mu}}_{NP}(\hat{\gamma}_{NR}) - \boldsymbol{\mu}^*) \xrightarrow{d} N(0, I)$$

where $\bar{\Omega}_{NR}$ is block-diagonal, with $(\bar{\Omega}_{1NR}, ..., \bar{\Omega}_{GNR})$ along the diagonal. Consider the following estimator of $\bar{\Omega}_{gNR}$:

$$\hat{\Omega}_{gNPR} = \frac{1}{NP}\sum_{t \in \mathcal{P}}\sum_{i=1}^{N}\hat{U}_{ig,NR}^2(\mathbf{Y}_{it} - \bar{\mathbf{Y}}_i)(\mathbf{Y}_{it} - \bar{\mathbf{Y}}_i)' = \frac{1}{NP}\sum_{t \in \mathcal{P}}\sum_{i=1}^{N}\hat{U}_{ig,NR}^2\hat{\boldsymbol{\varepsilon}}_{it}\hat{\boldsymbol{\varepsilon}}_{it}'$$

This can be shown to be consistent for $\bar{\Omega}_{gNR}$ using Kolmogorov's law of large numbers for *inid* random variables (e.g., Theorem 3.7 of White, 2001), and noting that Assumption 1(a) ensures the $2+\delta$ moment condition on $\boldsymbol{\varepsilon}_{it}$ and the finiteness of $\Sigma_i \, \forall i$. This holds for all $g$, and so we have $\hat{\Omega}_{NPR} - \bar{\Omega}_{NR} \xrightarrow{p} 0$. This implies that

$$\sqrt{NP}\hat{\Omega}_{NPR}^{-1/2}(\tilde{\boldsymbol{\mu}}_{NP}(\hat{\gamma}_{NR}) - \boldsymbol{\mu}^*) \xrightarrow{d} N(\mathbf{0}, I)$$

Under the null hypothesis of one cluster we have $\boldsymbol{\mu}^* = \boldsymbol{\iota}_G \otimes \boldsymbol{\mu}^\sharp$ where $\boldsymbol{\mu}^\sharp$ is some $(d \times 1)$ vector, implying that $A_{dG}\boldsymbol{\mu}^* = \mathbf{0}_{d(G-1)}$. Thus the $F$-statistic obeys

$$F_{NPR} = NP\tilde{\boldsymbol{\mu}}_{NP}'(\hat{\gamma}_{NR})A_{d,G}'\left(A_{d,G}\hat{\Omega}_{NPR}A_{d,G}'\right)^{-1}A_{d,G}\tilde{\boldsymbol{\mu}}_{NP}(\hat{\gamma}_{NR}) \xrightarrow{d} \chi^2_{d(G-1)}$$

As the limiting distribution of the $F$-statistic does not depend on $\mathcal{F}_R$, its unconditional distribution is also $\chi^2_{d(G-1)}$, completing the proof.

(b) Note that $\tilde{\boldsymbol{\mu}}_{NP}(\hat{\gamma}_{NR}) - \boldsymbol{\mu}^* = (\hat{\boldsymbol{\mu}}_{NR} - \boldsymbol{\mu}^*) + (\tilde{\boldsymbol{\mu}}_{NP}(\hat{\gamma}_{NR}) - \hat{\boldsymbol{\mu}}_{NR})$. Our Assumption 1 is sufficient for Assumption 1 of Bonhomme and Manresa (2015), and their Theorem 1 implies that the first term is $o_p(1)$ as $N, R \to \infty$. The second term is:

$$\begin{aligned}
\tilde{\boldsymbol{\mu}}_{g,NP}(\hat{\gamma}_{NR}) - \hat{\boldsymbol{\mu}}_{g,NR} &= \frac{1}{NP}\sum_{i=1}^{N}\sum_{t \in \mathcal{P}}\mathbf{Y}_{i,t}\hat{\pi}_{g,R}^{-1}\mathbf{1}\{\hat{\gamma}_{i,NR} = g\} - \frac{1}{NR}\sum_{i=1}^{N}\sum_{t \in \mathcal{R}}\mathbf{Y}_{i,t}\hat{\pi}_{g,R}^{-1}\mathbf{1}\{\hat{\gamma}_{i,NR} = g\} \\
&= \frac{1}{N}\sum_{i=1}^{N}\hat{\pi}_{g,R}^{-1}\mathbf{1}\{\hat{\gamma}_{i,NR} = g\}\left(\frac{1}{P}\sum_{t \in \mathcal{P}}\boldsymbol{\varepsilon}_{i,t} - \frac{1}{R}\sum_{t \in \mathcal{R}}\boldsymbol{\varepsilon}_{i,t}\right) \\
&\leq \underline{\pi}^{-1}\left(\frac{1}{NP}\sum_{i=1}^{N}\sum_{t \in \mathcal{P}}\boldsymbol{\varepsilon}_{i,t} - \frac{1}{NR}\sum_{i=1}^{N}\sum_{t \in \mathcal{R}}\boldsymbol{\varepsilon}_{i,t}\right) \\
&= \underline{\pi}^{-1}\left(O_p\left((NP)^{-1/2}\right) + O_p\left((NR)^{-1/2}\right)\right) \\
&= o_p(1), \text{ as } N, P, R \to \infty.
\end{aligned}$$

The penultimate line follows from a law of large numbers for *inid* data (e.g. Theorem 3.7 of White, 2001) the conditions for which are satisfied given our Assumption 1. This holds for $g = 1, ..., G$, and thus $\tilde{\boldsymbol{\mu}}_{NP}(\hat{\boldsymbol{\gamma}}_{NR}) \xrightarrow{p} \boldsymbol{\mu}^*$ as $N, P, R \to \infty$. This implies that

$$\tilde{\boldsymbol{\mu}}'_{NP}(\hat{\boldsymbol{\gamma}}_{NR}) A'_{d,G} \left(A_{d,G}\hat{\Omega}_{NPR}A'_{d,G}\right)^{-1} A_{d,G}\tilde{\boldsymbol{\mu}}_{NP}(\hat{\boldsymbol{\gamma}}_{NR}) \xrightarrow{p} \boldsymbol{\mu}^{*\prime} A'_{d,G}\left(A_{d,G}\bar{\Omega}_{NR}A'_{d,G}\right)^{-1} A_{d,G}\boldsymbol{\mu}^* > 0$$

by Assumption 2′(b) (clusters are "well separated"), the positive definiteness of $\bar{\Omega}_{NR}$, and the full row rank of $A_{d,G}$. Thus

$$F_{NPR} = NP\tilde{\boldsymbol{\mu}}'_{NP}(\hat{\boldsymbol{\gamma}}_{NR}) A'_{d,G}\left(A_{d,G}\hat{\Omega}_{NPR}A'_{d,G}\right)^{-1} A_{d,G}\tilde{\boldsymbol{\mu}}_{NP}(\hat{\boldsymbol{\gamma}}_{NR}) \xrightarrow{p} \infty, \text{ as } N, P, R \to \infty$$

completing the proof. ∎

Theorem 3 below requires a consistency result for clustering on a general estimated parameter vector, which we provide in Lemma 2, extending a result of Bonhomme and Manresa (2015).

**Lemma 2** *Under Assumption 4,* $\frac{1}{N}\sum_{i=1}^{N} \left\|\hat{\boldsymbol{\alpha}}_{NT,\hat{\gamma}_{NT,i}} - \boldsymbol{\alpha}^*_{\gamma^*_i}\right\|^2 \xrightarrow{p} 0$ *as* $N, T \to \infty$.

**Proof of Lemma 2.** We build on the proof of Theorem 1 of Bonhomme and Manresa (2015) (BM) for this result. To streamline notation in this proof, denote $\left(\hat{\boldsymbol{\beta}}_{NT}, \hat{\boldsymbol{\gamma}}_{NT}\right)$ as $\left(\hat{\boldsymbol{\beta}}, \hat{\boldsymbol{\gamma}}\right)$. All limits are for $N, T \to \infty$. The objective function we use in estimation is

$$\hat{Q}(\boldsymbol{\alpha}, \boldsymbol{\gamma}) \equiv \frac{1}{N}\sum_{i=1}^{N}\left\|\hat{\boldsymbol{\beta}}_{iT} - \boldsymbol{\alpha}_{\gamma_i}\right\|^2$$

Note that this objective function can be written as

$$\begin{aligned}\hat{Q}(\boldsymbol{\alpha}, \boldsymbol{\gamma}) &= \frac{1}{N}\sum_{i=1}^{N}\left\|\mathbf{Z}^*_{i,T}/\sqrt{T} + \left(\boldsymbol{\alpha}^*_{\gamma^*_i} - \boldsymbol{\alpha}_{\gamma_i}\right)\right\|^2, \text{ since } \boldsymbol{\beta}^*_i = \boldsymbol{\alpha}^*_{\gamma^*_i} \\ &= \frac{1}{N}\sum_{i=1}^{N}\left\|\boldsymbol{\alpha}^*_{\gamma^*_i} - \boldsymbol{\alpha}_{\gamma_i}\right\|^2 + \frac{1}{NT}\sum_{i=1}^{N}\left\|\mathbf{Z}^*_{i,T}\right\|^2 \\ &\quad + \frac{2}{N\sqrt{T}}\sum_{i=1}^{N}\left(\boldsymbol{\alpha}^*_{\gamma^*_i} - \boldsymbol{\alpha}_{\gamma_i}\right)' \mathbf{Z}^*_{i,T}\end{aligned}$$

Now consider the auxiliary objective function:

$$\tilde{Q}(\boldsymbol{\alpha}, \boldsymbol{\gamma}) \equiv \frac{1}{N}\sum_{i=1}^{N}\left\|\boldsymbol{\alpha}^*_{\gamma^*_i} - \boldsymbol{\alpha}_{\gamma_i}\right\|^2 + \frac{1}{NT}\sum_{i=1}^{N}\left\|\mathbf{Z}^*_{i,T}\right\|^2$$

Analogous to Lemma A.1 of BM, we now show that

$$p\lim_{N,T\to\infty} \sup_{\mathcal{A}^G \times \Gamma_G} \left| \hat{Q}(\boldsymbol{\alpha}, \boldsymbol{\gamma}) - \tilde{Q}(\boldsymbol{\alpha}, \boldsymbol{\gamma}) \right| = 0$$

Note that

$$\hat{Q}(\boldsymbol{\alpha}, \boldsymbol{\gamma}) - \tilde{Q}(\boldsymbol{\alpha}, \boldsymbol{\gamma}) = \frac{2}{N\sqrt{T}} \sum_{i=1}^{N} \left(\boldsymbol{\alpha}^*_{\gamma^*_i}\right)' \mathbf{Z}^*_{i,T} - \frac{2}{N\sqrt{T}} \sum_{i=1}^{N} \boldsymbol{\alpha}'_{\gamma_i} \mathbf{Z}^*_{i,T}$$

We next show that $\frac{1}{N} \sum_{i=1}^{N} \boldsymbol{\alpha}'_{\gamma_i} \mathbf{Z}^*_{i,T}$ is $o_p(1)$ uniformly on the parameter space.

$$\frac{1}{N\sqrt{T}} \sum_{i=1}^{N} \boldsymbol{\alpha}'_{\gamma_i} \mathbf{Z}^*_{i,T} = \frac{1}{N\sqrt{T}} \sum_{g=1}^{G} \sum_{i=1}^{N} \boldsymbol{\alpha}'_g \mathbf{Z}^*_{i,T} \mathbf{1}\{\gamma_i = g\}$$

$$= \sum_{g=1}^{G} \boldsymbol{\alpha}'_g \left( \frac{1}{N\sqrt{T}} \sum_{i=1}^{N} \mathbf{Z}^*_{i,T} \mathbf{1}\{\gamma_i = g\} \right)$$

$$\equiv \sum_{k=1}^{b} \sum_{g=1}^{G} \alpha_{k,g} \left( \frac{1}{N\sqrt{T}} \sum_{i=1}^{N} Z^*_{k,i,T} \mathbf{1}\{\gamma_i = g\} \right)$$

By the Cauchy-Schwarz inequality we have, for each $g \in \{1,...,G\}$:

$$\left( \sum_{k=1}^{b} \alpha_{k,g} \left( \frac{1}{N\sqrt{T}} \sum_{i=1}^{N} Z^*_{k,i,T} \mathbf{1}\{\gamma_i = g\} \right) \right)^2 \leq \left( \sum_{k=1}^{b} \alpha^2_{k,g} \right) \sum_{k=1}^{b} \left( \frac{1}{N\sqrt{T}} \sum_{i=1}^{N} Z^*_{k,i,T} \mathbf{1}\{\gamma_i = g\} \right)^2$$

Since $\boldsymbol{\alpha}_g$ is bounded $\sum_{k=1}^{b} \alpha^2_{k,g}$ is bounded. Then

$$\sum_{k=1}^{b} \left( \frac{1}{N\sqrt{T}} \sum_{i=1}^{N} Z^*_{k,i,T} \mathbf{1}\{\gamma_i = g\} \right)^2 = \frac{1}{N^2 T} \sum_{k=1}^{b} \sum_{i=1}^{N} \sum_{j=1}^{N} Z^*_{k,i,T} Z^*_{k,j,T} \mathbf{1}\{\gamma_i = g\} \mathbf{1}\{\gamma_j = g\}$$

$$\leq \frac{1}{N^2 T} \sum_{k=1}^{b} \sum_{i=1}^{N} \sum_{j=1}^{N} |Z_{k,i,T} Z_{k,j,T}|$$

$$+ \frac{1}{N^2 T} \sum_{k=1}^{b} \sum_{i=1}^{N} \sum_{j=1}^{N} (|Z_{k,i,T} \epsilon_{k,j,T}| + |\epsilon_{k,i,T} Z_{k,j,T}|)$$

$$+ \frac{1}{N^2 T} \sum_{k=1}^{b} \sum_{i=1}^{N} \sum_{j=1}^{N} |\epsilon_{k,i,T} \epsilon_{k,j,T}|$$

$$= \frac{1}{N^2 T} \sum_{k=1}^{b} \sum_{i=1}^{N} \sum_{j=1}^{N} |Z_{k,i,T} Z_{k,j,T}| + o_p\left(T^{-3/2}\right)$$

The first term on the RHS can be bounded using Cauchy-Schwarz:

$$\left( \frac{1}{N^2 T} \sum_{k=1}^{b} \sum_{i=1}^{N} \sum_{j=1}^{N} |Z_{k,i,T} Z_{k,j,T}| \right)^2 \leq \frac{1}{N^2 T} \sum_{k=1}^{b} \sum_{i=1}^{N} \sum_{j=1}^{N} Z^2_{k,i,T} Z^2_{k,j,T}$$

and this term is bounded in expectation:

$$\frac{1}{N^2 T} \sum_{k=1}^{b} \sum_{i=1}^{N} \sum_{j=1}^{N} E\left[Z^2_{k,i,T} Z^2_{k,j,T}\right] = \frac{1}{N^2 T} \sum_{k=1}^{b} \sum_{i=1}^{N} \sum_{j=1}^{N} E\left[Z^2_{k,i,T}\right] E\left[Z^2_{k,j,T}\right]$$

$$= \frac{1}{N^2 T} \sum_{k=1}^{b} \sum_{i=1}^{N} \sum_{j=1}^{N} V_{k,i} V_{k,j}$$

$$= O\left(T^{-1}\right)$$

This establishes that $\frac{1}{N\sqrt{T}}\sum_{i=1}^{N}\mathbf{Z}_{i,T}^{*\prime}\boldsymbol{\alpha}_{\gamma_i}$ is uniformly $o_p(1)$, and thus that $p\lim_{N,T\to\infty}\sup_{\mathcal{A}^G\times\Gamma_G}\left|\hat{Q}(\boldsymbol{\alpha},\boldsymbol{\gamma})-\tilde{Q}(\boldsymbol{\alpha},\boldsymbol{\gamma})\right|=0$. Next we show that $\tilde{Q}(\boldsymbol{\alpha},\boldsymbol{\gamma})$ is uniquely minimized at the true parameter values:

$$\begin{aligned}\tilde{Q}(\boldsymbol{\alpha},\boldsymbol{\gamma})-\tilde{Q}(\boldsymbol{\alpha}^*,\boldsymbol{\gamma}^*) &= \frac{1}{N}\sum_{i=1}^{N}\left\|\boldsymbol{\alpha}_{\gamma_i^*}^*-\boldsymbol{\alpha}_{\gamma_i}\right\|^2 + \frac{1}{NT}\sum_{i=1}^{N}\left\|\mathbf{Z}_{i,T}^*\right\|^2 \\ &\quad -\frac{1}{N}\sum_{i=1}^{N}\left\|\boldsymbol{\alpha}_{\gamma_i^*}^*-\boldsymbol{\alpha}_{\gamma_i^*}^*\right\|^2 - \frac{1}{NT}\sum_{i=1}^{N}\left\|\mathbf{Z}_{i,T}^*\right\|^2 \\ &= \frac{1}{N}\sum_{i=1}^{N}\left\|\boldsymbol{\alpha}_{\gamma_i^*}^*-\boldsymbol{\alpha}_{\gamma_i}\right\|^2 \\ &\geq 0\end{aligned}$$

with equality holding iff $\boldsymbol{\alpha}_{\gamma_i}=\boldsymbol{\alpha}_{\gamma_i^*}^*$. (This is analogous to Lemma A.2 of BM.) Combining the above results we obtain

$$\tilde{Q}(\hat{\boldsymbol{\alpha}},\hat{\boldsymbol{\gamma}})=\hat{Q}(\hat{\boldsymbol{\alpha}},\hat{\boldsymbol{\gamma}})+o_p(1)\leq\hat{Q}(\boldsymbol{\alpha}^*,\boldsymbol{\gamma}^*)+o_p(1)=\tilde{Q}(\boldsymbol{\alpha}^*,\boldsymbol{\gamma}^*)+o_p(1)$$

This implies that $\tilde{Q}(\hat{\boldsymbol{\alpha}},\hat{\boldsymbol{\gamma}})-\tilde{Q}(\boldsymbol{\alpha}^*,\boldsymbol{\gamma}^*)=o_p(1)$, and we note that

$$\tilde{Q}(\hat{\boldsymbol{\alpha}},\hat{\boldsymbol{\gamma}})-\tilde{Q}(\boldsymbol{\alpha}^*,\boldsymbol{\gamma}^*)=\frac{1}{N}\sum_{i=1}^{N}\left\|\boldsymbol{\alpha}_{\gamma_i^*}^*-\hat{\boldsymbol{\alpha}}_{\hat{\gamma}_i}\right\|^2$$

completing the proof. ∎

**Proof of Theorem 3.**  (a) We first find the limiting distribution of $\sqrt{NP}\tilde{\boldsymbol{\alpha}}_{NP}(\hat{\boldsymbol{\gamma}}_{NR})$ conditional on $\mathcal{F}_R$. Note that

$$\begin{aligned}\tilde{\boldsymbol{\alpha}}_{g,NP}(\hat{\boldsymbol{\gamma}}_{NR})-\boldsymbol{\alpha}_g^* &= \frac{1}{N}\sum_{i=1}^{N}\hat{\boldsymbol{\beta}}_{i,P}\hat{\pi}_{g,NR}^{-1}\mathbf{1}\{\hat{\gamma}_{i,NR}=g\} \\ &= \frac{1}{N\sqrt{P}}\sum_{i=1}^{N}\mathbf{Z}_{i,P}\hat{\pi}_{g,NR}^{-1}\mathbf{1}\{\hat{\gamma}_{i,NR}=g\}+\frac{1}{N\sqrt{P}}\sum_{i=1}^{N}\boldsymbol{\epsilon}_{i,P}\hat{\pi}_{g,NR}^{-1}\mathbf{1}\{\hat{\gamma}_{i,NR}=g\}\end{aligned}$$

The second term on the RHS is bounded by

$$\frac{1}{N\sqrt{P}}\sum_{i=1}^{N}\boldsymbol{\epsilon}_{i,P}\hat{\pi}_{g,NR}^{-1}\mathbf{1}\{\hat{\gamma}_{i,NR}=g\}\leq\underline{\pi}^{-1}\frac{1}{N\sqrt{P}}\sum_{i=1}^{N}\boldsymbol{\epsilon}_{i,P}=o_p\left((NP)^{-1/2}\right)$$

by Assumption 4(d). Similar to Theorem 1, we then have

$$\sqrt{NP}\left(\tilde{\boldsymbol{\alpha}}_{g,NP}(\hat{\boldsymbol{\gamma}}_{NR})-\boldsymbol{\alpha}_g^*\right)=\frac{1}{\sqrt{N}}\sum_{i=1}^{N}\hat{U}_{ig,NR}\mathbf{Z}_{i,P}+o_p(1)$$

where $\hat{U}_{ig,NR} \equiv \hat{\pi}_{g,R}^{-1} \mathbf{1}\{\hat{\gamma}_{i,NR} = g\}$. We obtain following limiting distribution

$$\sqrt{NP}\bar{\Omega}_{NR}^{-1/2}(\tilde{\boldsymbol{\alpha}}_{NP}(\hat{\boldsymbol{\gamma}}_{NR}) - \boldsymbol{\alpha}^*) \xrightarrow{d} N(\mathbf{0}, I)$$

using steps similar to those in Theorem 1(a), omitted in the interests of space. $\bar{\Omega}_{NR}$ is block-diagonal, with $(\bar{\Omega}_{1,NR}, ..., \bar{\Omega}_{G,NR})$ along the diagonal. By Assumption 4(b) we have $\hat{V}_{i,P} \xrightarrow{p} V_i$ for each $i$, so consider the estimator:

$$\hat{\Omega}_{g,NPR} = \frac{1}{N} \sum_{i=1}^{N} \hat{U}_{ig,NR}^2 \hat{V}_{i,P}$$

and note that $\hat{\Omega}_{g,NPR} - \bar{\Omega}_{g,NR} = \frac{1}{N}\sum_{i=1}^{N} \hat{U}_{ig,NR}^2 (\hat{V}_{i,P} - V_i) = o_p(1)$ by the consistency of $\hat{V}_{i,P}$ and the boundedness of $\hat{U}_{ig,NR}$. This holds for each $g$, and so we have $\hat{\Omega}_{NPR} - \bar{\Omega}_{NR} \xrightarrow{p} 0$. Thus

$$\sqrt{NP}\hat{\Omega}_{NPR}^{-1/2}(\tilde{\boldsymbol{\alpha}}_{NP}(\hat{\boldsymbol{\gamma}}_{NR}) - \boldsymbol{\alpha}^*) \xrightarrow{d} N(\mathbf{0}, I)$$

Under Assumption $2_P$ we have $\boldsymbol{\alpha}^* = \boldsymbol{\iota}_G \otimes \boldsymbol{\alpha}^\sharp$ where $\boldsymbol{\alpha}^\sharp$ is some $(d \times 1)$ vector, which implies that $A_{bG}\boldsymbol{\alpha}^* = \mathbf{0}_{b(G-1)}$. Thus the test statistic obeys

$$F_{NPR} = NP\tilde{\boldsymbol{\alpha}}'_{NP}(\hat{\boldsymbol{\gamma}}_{NR}) A'_{b,G} \left(A_{b,G}\hat{\Omega}_{NPR}A'_{b,G}\right)^{-1} A_{b,G}\tilde{\boldsymbol{\alpha}}_{NP}(\hat{\boldsymbol{\gamma}}_{NR}) \xrightarrow{d} \chi^2_{b(G-1)}$$

As the limiting distribution of the test statistic does not depend on $\mathcal{F}_R$, its unconditional distribution is also $\chi^2_{b(G-1)}$, completing this part of the proof.

(b) Note that $\tilde{\boldsymbol{\alpha}}_{NP}(\hat{\boldsymbol{\gamma}}_{NR}) - \boldsymbol{\alpha}^* = (\hat{\boldsymbol{\alpha}}_{NR} - \boldsymbol{\alpha}^*) + (\tilde{\boldsymbol{\alpha}}_{NP}(\hat{\boldsymbol{\gamma}}_{NR}) - \hat{\boldsymbol{\alpha}}_{NR})$. Lemma 2 implies that the first term on the RHS is $o_p(1)$ as $N, R \to \infty$, and derivations very similar to those in the proof of Theorem 1(b) show that $\tilde{\boldsymbol{\alpha}}_{g,NP}(\hat{\boldsymbol{\gamma}}_{NR}) - \hat{\boldsymbol{\alpha}}_{g,NR} = o_p(1)$. These hold for $g = 1, ..., G$, and thus $\tilde{\boldsymbol{\alpha}}_{NP}(\hat{\boldsymbol{\gamma}}_{NR}) \xrightarrow{p} \boldsymbol{\alpha}^*$, as $N, P, R \to \infty$. This implies that

$$\tilde{\boldsymbol{\alpha}}'_{NP}(\hat{\boldsymbol{\gamma}}_{NR}) A'_{b,G}\left(A_{b,G}\hat{\Omega}_{NPR}A'_{b,G}\right)^{-1} A_{b,G}\tilde{\boldsymbol{\alpha}}_{NP}(\hat{\boldsymbol{\gamma}}_{NR}) \xrightarrow{p} \boldsymbol{\alpha}^{*\prime}A'_{b,G}\left(A_{b,G}\Omega_{NR}A'_{b,G}\right)^{-1} A_{b,G}\boldsymbol{\alpha}^* > 0$$

by Assumption $2'_P$(b), the positive definiteness of $\bar{\Omega}_{NR}$, and the full row rank of $A_{b,G}$. Thus $F_{NPR} \xrightarrow{p} \infty$ as $N, P, R \to \infty$. ∎

**Proof of Theorem 4.** (a) This is done using the same methods as Theorem 1 for $d = 1$ and $G = \hat{G}_{NR}$. We note that the re-ordering of the clusters (from largest to smallest) is known given $\mathcal{F}_R$, as is the value of $\hat{G}_{NR}$. Thus, following the steps in the proof of Theorem 1(a) we have $F_{NPR} \xrightarrow{d} \chi_q^2$ where $q = \hat{G}_{NR} - 1$. This limit distribution depends on $\mathcal{F}_R$ via the value of $\hat{G}_{NR}$; by transforming the test statistic using its limiting CDF we obtain $1 - Pval_{NPR} \xrightarrow{d} Unif(0,1) \Rightarrow Pval_{NPR} \xrightarrow{d} Unif(0,1)$, both conditional on $\mathcal{F}_R$, and since the limit distribution does not depend on $\mathcal{F}_R$, this result also holds unconditionally.

(b) As in the proof of Theorem 1(b), note that $\tilde{\boldsymbol{\mu}}_{NP}(\hat{\boldsymbol{\gamma}}_{NR}) - \boldsymbol{\mu}^* = (\hat{\boldsymbol{\mu}}_{NR} - \boldsymbol{\mu}^*) + (\tilde{\boldsymbol{\mu}}_{NP}(\hat{\boldsymbol{\gamma}}_{NR}) - \hat{\boldsymbol{\mu}}_{NR})$. Our Assumption 1 is sufficient for Assumption 1 of Bonhomme and Manresa (2015), and their Theorem 1 implies that the first term on the RHS is $o_p(1)$, as $N, R \to \infty$. (Note that their Theorem 1 does not require non-negligibility of group sizes.) The first $\hat{G}_{NR}$ elements of the second term are $o_p(1)$ as $N, P, R \to \infty$ using the same derivation as in the proof of Theorem 1(b), noting that the condition that $\underline{\pi} > 0$ holds for $g \in \{1, ..., \hat{G}_{NR}\}$, and we have $\hat{G}_{NR} \geq 2$ by Assumption $2'(c^S)$. This implies that

$$\tilde{\boldsymbol{\mu}}'_{NP}(\hat{\boldsymbol{\gamma}}_{NR}) B'_{\hat{G}_{NR},G} \left( B_{\hat{G}_{NR},G} \hat{\Omega}_{NPR} B'_{\hat{G}_{NR},G} \right)^{-1} B_{\hat{G}_{NR},G} \tilde{\boldsymbol{\mu}}_{NP}(\hat{\boldsymbol{\gamma}}_{NR})$$
$$\to \boldsymbol{\mu}^{*\prime} B'_{\hat{G}_{NR},G} \left( B_{\hat{G}_{NR},G} \hat{\Omega}_{NPR} B'_{\hat{G}_{NR},G} \right)^{-1} B_{\hat{G}_{NR},G} \boldsymbol{\mu}^* > 0$$

by Assumption $2'(b)$, the positive definiteness of $\bar{\Omega}_{NP}$, and the full row rank of $B_{1,\hat{G}_{NR},G}$. Thus $F_{NPR} \xrightarrow{p} \infty$ and $P_{NPR} \xrightarrow{p} 0$ as $N, P, R \to \infty$. ∎

**Proof of Theorem 5.** (a) Note

$$\sqrt{NP}\left(\tilde{\mu}_{1,NP}(\hat{\gamma}_{NR}) - \tilde{\mu}_{2,NP}(\hat{\gamma}_{NR})\right) = \frac{1}{\sqrt{NP}} \sum_{i=1}^{N} \sum_{t \in \mathcal{P}} \varepsilon_{it} \left(\hat{\pi}_{1,R}^{-1} \mathbf{1}\{\hat{\gamma}_{i,NR} = 1\} - \hat{\pi}_{2,R}^{-1} \mathbf{1}\{\hat{\gamma}_{i,NR} = 2\}\right)$$
$$= \frac{1}{\sqrt{NP}} \sum_{i=1}^{N} \sum_{t \in \mathcal{P}} \hat{Z}_{i,NR} \varepsilon_{it}$$

where $\hat{Z}_{i,NR} \equiv \hat{\pi}_{1,R}^{-1} \mathbf{1}\{\hat{\gamma}_{i,NR} = 1\} - \hat{\pi}_{2,R}^{-1} \mathbf{1}\{\hat{\gamma}_{i,NR} = 2\}$. Conditional on $\mathcal{F}_R$, the weights, $\hat{Z}_{i,NR}$, on $\varepsilon_{it}$ are known, and they are bounded since $\underline{\pi} > 0$. Define the variable $\xi_{it,NR} \equiv \hat{Z}_{i,NR} \varepsilon_{it}$, and note that conditional on $\mathcal{F}_R$, $\xi_{it,NR}$ is independent of $\xi_{jt,NR}$.

Denote the observations $t \in \mathcal{P}$ as $(t_1, ..., t_P)$, and define

$$\boldsymbol{\xi}'_{i,NPR} \equiv [\xi_{i,t_1,NR}, ..., \xi_{i,t_P,NR}] = \hat{Z}_{i,NR} [\varepsilon_{i,t_1}, ..., \varepsilon_{i,t_P}]' \equiv \hat{Z}_{i,NR} \boldsymbol{\varepsilon}'_{iP}$$

and note that $E\left[\boldsymbol{\xi}_{i,NPR}\boldsymbol{\xi}'_{i,NPR}|\mathcal{F}_R\right] = \hat{Z}^2_{i,NR}E\left[\boldsymbol{\varepsilon}_{iP}\boldsymbol{\varepsilon}'_{iP}\right] \equiv \hat{Z}^2_{i,NR}\Omega_i$. Finally, define

$$\bar{\omega}^2_{NR} \equiv \lim_{N,P\to\infty} \frac{1}{N}\sum_{i=1}^N \hat{Z}^2_{i,NR}\boldsymbol{\iota}'_P\Omega_i\boldsymbol{\iota}_P$$

$$\text{and} \quad \hat{\omega}^2_{NPR} \equiv \frac{1}{NP}\sum_{i=1}^N \hat{Z}^2_{i,NR}\boldsymbol{\iota}'_P\hat{\boldsymbol{\varepsilon}}_i\hat{\boldsymbol{\varepsilon}}'_i\boldsymbol{\iota}_P$$

where $\hat{\boldsymbol{\varepsilon}}_{iP} = \mathbf{Y}_{iP} - \tilde{\mu}_{\hat{\gamma}_{i,NR},NP}(\hat{\gamma}_{NR})$ and $\mathbf{Y}_{iP} \equiv \{Y_{it}\}_{t\in\mathcal{P}}$. Our Assumption 1′ is sufficient for Assumptions 1, 2 and 3(b) of Hansen (2007, Theorem 1), and thus we have, conditional on $\mathcal{F}_R$,

$$\frac{\sqrt{NP}\frac{1}{NP}\sum_{i=1}^N\sum_{t\in\mathcal{P}}\hat{Z}_{i,NR}\varepsilon_{it}}{\bar{\omega}_{NR}} \xrightarrow{d} N(0,1) \quad \text{and} \quad \hat{\omega}^2_{NPR} \xrightarrow{p} \bar{\omega}^2_{NR}, \text{ as } N\to\infty$$

This implies that the $t$-statistic obeys

$$\frac{\sqrt{NP}\frac{1}{NP}\sum_{i=1}^N\sum_{t\in\mathcal{P}}\hat{Z}_{i,NR}\varepsilon_{it}}{\hat{\omega}_{NPR}} \xrightarrow{d} N(0,1), \text{ as } N\to\infty$$

As the limiting distribution of the $t$-statistic does not depend on $\mathcal{F}_R$, its unconditional distribution is also $N(0,1)$, completing the proof.

(b) As in Theorem 1(b), note that $\tilde{\boldsymbol{\mu}}_{NP}(\hat{\boldsymbol{\gamma}}_{NR}) - \boldsymbol{\mu}^* = (\hat{\boldsymbol{\mu}}_{NR} - \boldsymbol{\mu}^*) + (\tilde{\boldsymbol{\mu}}_{NP}(\hat{\boldsymbol{\gamma}}_{NR}) - \hat{\boldsymbol{\mu}}_{NR})$. Our Assumption 1′ is sufficient for assumptions in Bonhomme and Manresa (2015, Appendix S2.2.1), which implies that the first term on the RHS is $o_p(1)$, as $N\to\infty$. The second term is:

$$\begin{aligned}
\tilde{\mu}_{g,NP}(\hat{\boldsymbol{\gamma}}_{NR}) - \hat{\mu}_{g,NR} &= \frac{1}{N}\sum_{i=1}^N \hat{\pi}^{-1}_{g,R}\mathbf{1}\{\hat{\gamma}_{i,NR}=g\}\left(\frac{1}{P}\sum_{t\in\mathcal{P}}Y_{i,t} - \frac{1}{R}\sum_{t\in\mathcal{R}}Y_{i,t}\right) \\
&= \frac{1}{N}\sum_{i=1}^N \hat{\pi}^{-1}_{g,R}\mathbf{1}\{\hat{\gamma}_{i,NR}=g\}\left(\frac{1}{P}\sum_{t\in\mathcal{P}}\varepsilon_{i,t} - \frac{1}{R}\sum_{t\in\mathcal{R}}\varepsilon_{i,t}\right) \\
&\leq \underline{\pi}^{-1}\left(\frac{1}{NP}\sum_{i=1}^N\sum_{t\in\mathcal{P}}\varepsilon_{i,t} - \frac{1}{NR}\sum_{i=1}^N\sum_{t\in\mathcal{R}}\varepsilon_{i,t}\right) \\
&= o_p(1), \text{ as } N\to\infty
\end{aligned}$$

since our Assumption 1′ is sufficient for Assumptions 1, 2 and 3(b) of Hansen (2007, Theorem 1). This holds for $g=1,2$, and thus $\tilde{\boldsymbol{\mu}}_{NP}(\hat{\boldsymbol{\gamma}}_{NR}) \xrightarrow{p} \boldsymbol{\mu}^*$, as $N\to\infty$. This implies $\tilde{\mu}_{1,NP}(\hat{\boldsymbol{\gamma}}_{NR}) - \tilde{\mu}_{2,NP}(\hat{\boldsymbol{\gamma}}_{NR}) \xrightarrow{p} \mu^*_1 - \mu^*_2 \neq 0$ by Assumption 2′(b). Thus $|tstat_{NPR}| \xrightarrow{p} \infty$, as $N\to\infty$. ∎

# References

[1] Arthur, D. and S. Vassilvitskii, 2007, K-means++: The advantages of careful seeding, in *Proceedings of the Eighteenth Annual ACM-SIAM Symposium on Discrete Algorithms, SODA '07*, Philadelphia, PA, USA: Society for Industrial and Applied Mathematics, 1027–1035.

[2] Bonhomme, S., and Manresa, E., 2015, Grouped patterns of heterogeneity in panel data, *Econometrica*, 83(3), 1147-1184.

[3] Brown, S.J. and W.N. Goetzmann, 1997, Mutual fund styles, *Journal of Financial Economics*, 43 (3), 373-399.

[4] Carhart, M.M., 1997, On persistence in mutual fund performance, *Journal of Finance*, 52 (1), 57-82.

[5] Eisen, M.B., P.T. Spellman, P.O Brown, and D. Botstein, D., 1998, Cluster analysis and display of genome-wide expression patterns, *Proceedings of the National Academy Science*, 95(25), 14863-14868.

[6] Fraley, C. and Raftery, A.E., 2002, Model-based clustering, discriminant analysis, and density estimation, *Journal of the American Statistical Association*, 97 (458), 611-631.

[7] Francis, N., M.T. Owyang, and Ö. Savascin, 2017, An endogenously clustered factor approach to international business cycles, *Journal of Applied Econometrics*, 32, 1261-1276.

[8] Fu, W. and P.O. Perry, 2017, Estimating the number of clusters using cross-validation, working paper, Stern School of Business, New York University.

[9] García-Escudero, L.A. and A. Gordaliza, 1999, Robustness properties of $k$ means and trimed $k$ means, *Journal of the American Statistical Association*, 94(447), 956-969.

[10] Hansen, C.B., 2007, Asymptotic properties of a robust variance matrix estimator for panel data when $T$ is large, *Journal of Econometrics*, 141, 597-620.

[11] Liu, Y., Hayes, D.N., Nobel, A. and Marron, J.S., 2008, Statistical significance of clustering for high-dimension, low-sample size data, *Journal of the American Statistical Association*, 103 (483), 1281-1293.

[12] Maitra, R., Melnykov, V. and Lahiri, S.N., 2012, Bootstrapping for significance of compact clusters in multidimensional datasets, *Journal of the American Statistical Association*, 107 (497), 378-392.

[13] Ng, A.Y., M.I. Jordan and Y. Weiss, 2002, On spectral clustering: Analysis and an algorithm, in *Advances in neural information processing systems*, 849-856.

[14] Patton, A.J. and Weller, B.M., 2019, What you see is not what you get: The costs of trading market anomalies, *Journal of Financial Economics*, forthcoming.

[15] Pollard, D., 1981, Strong consistency of $k$-means clustering, *Annals of Statistics*, 9, 135-140.

[16] Pollard, D., 1982, A central limit theorem for $k$-means clustering, *Annals of Probability*, 10, 919-926.

[17] Ray, S., and R.H. Turi, 1999, Determination of number of clusters in k-means clustering and application in colour image segmentation, in *Proceedings of the 4th International Conference on Advances in Pattern Recognition and Digital Techniques*, 137-143.

[18] Steinbach, M., G. Karypis, and V. Kumar, 2000, A comparison of document clustering techniques, in *KDD Workshop on Text Mining*, 400(1), 525-526.

[19] Sugar, C.A. and James, G.M., 2003, Finding the number of clusters in a dataset, *Journal of the American Statistical Association*, 98 (463), 750-763.

[20] Tibshirani, R., Walther, G. and Hastie, T.J., 2003, Estimating the number of clusters in a dataset via the gap statistic, *Journal of the Royal Statistical Society,* Series B, 63, 411-423.

[21] Tibshirani, R. and Walther, G., 2005, Cluster validation by prediction strength, *Journal of Computational and Graphic Statistics*, 14(3), 511-528.

[22] Wang, J., 2010, Consistent selection of the number of clusters via crossvalidation, *Biometrika*, 97(4), 893-904.

[23] Witten, D.M. and R. Tibshirani, 2010, A framework for feature selection in clustering, *Journal of the American Statistical Association*, 105(490), 713-726.

Table 1: Finite sample rejection rates

| d | G | N = 30, T = 50 | 30, 250 | 30, 1000 | 150, 50 | 150, 250 | 150, 1000 | 600, 50 | 600, 250 | 600, 1000 |
|---|---|---|---|---|---|---|---|---|---|---|
| | | *Normal data* | | | | | | | | |
| 1 | 2 | 0.055 | 0.049 | 0.056 | 0.057 | 0.046 | 0.055 | 0.059 | 0.056 | 0.048 |
| 1 | 3 | 0.052 | 0.047 | 0.035 | 0.046 | 0.045 | 0.047 | 0.050 | 0.043 | 0.055 |
| 1 | 4 | 0.057 | 0.042 | 0.064 | 0.034 | 0.059 | 0.052 | 0.053 | 0.056 | 0.036 |
| 1 | 5 | 0.058 | 0.054 | 0.059 | 0.034 | 0.048 | 0.053 | 0.049 | 0.051 | 0.047 |
| 1 | Bonf. | 0.057 | 0.057 | 0.055 | 0.045 | 0.048 | 0.049 | 0.059 | 0.045 | 0.045 |
| 2 | 2 | 0.040 | 0.048 | 0.064 | 0.040 | 0.046 | 0.052 | 0.049 | 0.046 | 0.039 |
| 2 | 3 | 0.048 | 0.051 | 0.060 | 0.040 | 0.062 | 0.045 | 0.036 | 0.058 | 0.061 |
| 2 | 4 | 0.072 | 0.061 | 0.046 | 0.050 | 0.040 | 0.047 | 0.039 | 0.061 | 0.044 |
| 2 | 5 | 0.058 | 0.044 | 0.043 | 0.045 | 0.063 | 0.043 | 0.067 | 0.053 | 0.054 |
| 2 | Bonf. | 0.049 | 0.044 | 0.066 | 0.042 | 0.050 | 0.040 | 0.045 | 0.064 | 0.047 |
| 5 | 2 | 0.040 | 0.058 | 0.060 | 0.049 | 0.051 | 0.062 | 0.052 | 0.041 | 0.044 |
| 5 | 3 | 0.050 | 0.052 | 0.059 | 0.052 | 0.054 | 0.044 | 0.052 | 0.049 | 0.053 |
| 5 | 4 | 0.066 | 0.047 | 0.054 | 0.041 | 0.067 | 0.053 | 0.052 | 0.055 | 0.051 |
| 5 | 5 | 0.083 | 0.049 | 0.049 | 0.055 | 0.037 | 0.044 | 0.059 | 0.049 | 0.048 |
| 5 | Bonf. | 0.065 | 0.051 | 0.063 | 0.043 | 0.051 | 0.049 | 0.050 | 0.044 | 0.050 |
| | | *Heterogeneous data* | | | | | | | | |
| 1 | 2 | 0.055 | 0.045 | 0.040 | 0.042 | 0.061 | 0.060 | 0.049 | 0.040 | 0.055 |
| 1 | 3 | 0.062 | 0.046 | 0.062 | 0.041 | 0.057 | 0.050 | 0.047 | 0.047 | 0.057 |
| 1 | 4 | 0.045 | 0.045 | 0.053 | 0.053 | 0.062 | 0.053 | 0.052 | 0.052 | 0.036 |
| 1 | 5 | 0.058 | 0.050 | 0.057 | 0.052 | 0.053 | 0.039 | 0.051 | 0.044 | 0.057 |
| 1 | Bonf. | 0.060 | 0.043 | 0.053 | 0.040 | 0.052 | 0.056 | 0.050 | 0.045 | 0.051 |
| 2 | 2 | 0.048 | 0.049 | 0.048 | 0.045 | 0.048 | 0.045 | 0.042 | 0.053 | 0.043 |
| 2 | 3 | 0.053 | 0.039 | 0.045 | 0.050 | 0.046 | 0.062 | 0.049 | 0.045 | 0.052 |
| 2 | 4 | 0.063 | 0.059 | 0.051 | 0.055 | 0.053 | 0.050 | 0.037 | 0.045 | 0.058 |
| 2 | 5 | 0.049 | 0.037 | 0.075 | 0.053 | 0.038 | 0.041 | 0.041 | 0.047 | 0.041 |
| 2 | Bonf. | 0.050 | 0.044 | 0.049 | 0.053 | 0.039 | 0.046 | 0.038 | 0.043 | 0.049 |
| 5 | 2 | 0.054 | 0.049 | 0.044 | 0.048 | 0.049 | 0.057 | 0.044 | 0.047 | 0.039 |
| 5 | 3 | 0.056 | 0.038 | 0.050 | 0.053 | 0.045 | 0.055 | 0.052 | 0.043 | 0.040 |
| 5 | 4 | 0.076 | 0.067 | 0.039 | 0.063 | 0.047 | 0.048 | 0.053 | 0.045 | 0.054 |
| 5 | 5 | 0.069 | 0.055 | 0.050 | 0.070 | 0.041 | 0.047 | 0.040 | 0.057 | 0.051 |
| 5 | Bonf. | 0.066 | 0.054 | 0.050 | 0.047 | 0.032 | 0.062 | 0.040 | 0.046 | 0.046 |

Notes: This table presents the proportion of simulations in which we reject the null of a single cluster in favor of multiple clusters, using the test proposed in Theorem 1 at a 0.05 significance level. The top panel presents results for *iid* Normal data; the lower panel presents results when the distribution is randomly drawn from one of $N(0,1)$, $Exp(2)$, $Unif(-3,3)$, $\chi^2(4)$ or $t(5)$, each standardized to have zero mean and unit variance. The dimension of the variables is denoted $d$, the number of groups considered under the alternative is denoted $G$, the number of variables is denoted $N$, and the number of time series observations is denoted $T$. Rows labeled "Bonf." use tests with a Bonferroni correction to consider $G \in \{2,3,4,5\}$ under the alternative. The number of simulations is 1000.

## Table 2 Vehicle manufacturer clusters

**Panel A: Cluster properties by characteristic**

|         | **Acceleration** (s to 60mph) | **Cylinders** (#) | **Displacement** (in$^3$) | **Horsepower** (hp) | **MPG** (mpg) | **Weight** (log lbs) |
|---------|---------|---------|---------|---------|---------|---------|
| \multicolumn{7}{l}{Normalized values, $\mathcal{R}$ sample} | | | | | | |
| *Group 1* | -0.441 | 0.641  | 0.709  | 0.690  | -0.639 | 0.636  |
| *Group 2* | 0.326  | -1.110 | -1.057 | -0.704 | 0.973  | -0.949 |
| Normalized values, $\mathcal{P}$ sample | | | | | | |
| *Group 1* | -0.125 | 0.507  | 0.628  | 0.450  | -0.486 | 0.573  |
| *Group 2* | 0.152  | -0.557 | -0.650 | -0.318 | 0.266  | -0.461 |
| Raw values, $\mathcal{P}$ sample | | | | | | |
| *Group 1* | 15.76  | 5.82   | 217.25 | 105.02 | 23.21  | 8.04   |
| *Group 2* | 16.46  | 4.21   | 111.06 | 83.67  | 28.94  | 7.78   |

**Panel B: Cluster assignments**

| | | | | | | |
|---|---|---|---|---|---|---|
| *Group 1* | AMC     | Buick      | Chevrolet | Chrysler | Dodge  | Ford   |
|           | Mercury | Oldsmobile | Plymouth  | Pontiac  |        |        |
| *Group 2* | Audi        | BMW     | Datsun  | Fiat | Honda  | Mazda  |
|           | Opel        | Peugeot | Renault | Saab | Subaru | Toyota |
|           | Volkswagon  | Volvo   |         |      |        |        |

Notes: This table presents group averages of manufacturer-level characteristics in a $G = 2$ cluster model for $\mathcal{R}$ (1970–1975) and $\mathcal{P}$ (1976–1982) samples (Panel A) and manufacturer names by group (Panel B). The "raw values" in Panel A are renormalized using the $\mathcal{P}$-sample characteristic means and standard deviations.

Table 3: Mutual fund clusters

| Group | $\hat{N}_g$ | $\hat{\beta}_{MKT}$ | $\hat{\beta}_{SMB}$ | $\hat{\beta}_{HML}$ | $\hat{\beta}_{UMD}$ | $\hat{\alpha}$ | $\hat{\sigma}$ |
|---|---|---|---|---|---|---|---|
| | $\mathcal{R}$ sample | | | | | | |
| 1 | 420 | 0.905 | -0.088 | 0.073 | -0.022 | -0.817 | 4.579 |
| 2 | 198 | 1.266 | 0.925 | 0.287 | 0.233 | 17.875 | 11.263 |
| 3 | 84 | 1.320 | 0.677 | -0.342 | 0.414 | 21.722 | 15.294 |
| 4 | 238 | 1.058 | 0.685 | 0.705 | -0.164 | 3.500 | 9.224 |
| 5 | 224 | 0.840 | 0.368 | 0.341 | -0.026 | 5.463 | 8.469 |
| 6 | 210 | 0.959 | -0.017 | -0.286 | 0.176 | 0.216 | 7.561 |
| 7 | 270 | 1.004 | -0.004 | 0.522 | -0.194 | 0.768 | 6.791 |
| 8 | 99 | 0.029 | 0.047 | 0.075 | 0.006 | -3.155 | 4.994 |
| | $\mathcal{P}$ sample | | | | | | |
| 1 | 420 | 0.883 | -0.142 | 0.104 | -0.021 | 3.370 | 6.238 |
| 2 | 198 | 1.210 | 0.807 | -0.041 | 0.080 | 14.896 | 13.849 |
| 3 | 84 | 1.219 | 0.428 | -0.825 | 0.228 | 27.074 | 19.375 |
| 4 | 238 | 0.975 | 0.494 | 0.493 | -0.168 | 12.257 | 11.096 |
| 5 | 224 | 0.809 | 0.288 | 0.198 | -0.067 | 7.883 | 10.617 |
| 6 | 210 | 0.966 | -0.061 | -0.306 | 0.118 | 12.108 | 11.034 |
| 7 | 270 | 0.954 | -0.077 | 0.538 | -0.175 | 6.725 | 9.31 |
| 8 | 99 | 0.042 | 0.066 | 0.029 | -0.027 | 2.398 | 6.769 |

Notes: This table presents group averages of fund-level characteristics in a $G = 8$ group model for $\mathcal{R}$ (year 1999) and $\mathcal{P}$ (year 2000) samples. Average abnormal returns ($\alpha$) and idiosyncratic volatility ($\sigma$) are annualized and reported in percent.

Figure 1: *This figure shows test rejection frequencies as a function of $\mu_2$, holding $\mu_1 = 0$, for six different combinations of sample sizes, and for two types of data (Normally and heterogeneously distributed). When $\mu_2 = 0$ the rejection frequency should equal 0.05, the nominal size of the test.*

Figure 2: *This figure shows test rejection frequencies as a function of $\tilde{G}$, the number of clusters considered under the alternative, for two different combinations of sample sizes. In the left panel the correct number of clusters is 2, while in the right panel it is 5. The distance between the first and last (ordered) cluster means is 0.2 when $(N,T) = (30, 50)$ and 0.075 when $(N,T) = (150, 250)$.*

Figure 3: *This figure shows test rejection frequencies as a function of $\mu_2$, holding $\mu_1 = 0$, for two different combinations of sample sizes, and for two tests: the first uses $G = 2$ under the alternative, the second considers $G \in \{2, 3, 4, 5\}$ and uses a Bonferroni correction to control for multiple testing. When $\mu_2 = 0$ the rejection frequency should equal 0.05, the nominal size of the test.*

Figure 4: *This figure shows test rejection frequencies as a function of $\pi_3$, the fraction of variables in the smallest cluster. The fractions of variables in the other two groups is set at $(1 - \pi_3)/2$. Two different combinations of sample sizes are considered. The distance between the first and third (ordered) cluster means is zero in the left panel, and is 0.2 when $(N, T) = (30, 50)$ and 0.075 when $(N, T) = (150, 250)$ in the right panel. The nominal size of the test is 0.05.*

Figure 5: *This figure shows test rejection frequencies as a function of $\phi_2$, the AR(1) parameter of the second cluster, when the parameter for the first cluster is $\phi_1 = 0.5$. Two different combinations of sample sizes are considered. When $\phi_2 = 0.5$ the rejection frequency should equal 0.05, the nominal size of the test.*

Figure 6: *This figure shows test rejection frequencies as a function of $\mu_2$, holding $\mu_1 = 0$, for four different time series sample sizes, and three different cross-sectional sample sizes. When $\mu_2 = 0$ the rejection frequency should equal 0.05, the nominal size of the test.*

Supplemental Appendix for

# Testing for Unobserved Heterogeneity via *k-means* Clustering

by Andrew J. Patton and Brian M. Weller

15 July, 2019

## S.A.1: Extensions of Theorem 1

To streamline exposition, in this appendix we focus on the case that $d \equiv \dim(Y_{it}) = 1$ and $G = 2$. All of the results below hold for any finite value of $d$ and $G$.

We present a simplified version of Theorem 1 for $d = 1$, $G = 2$. In this instance, it is more natural to consider a $t$-test of the difference in cluster means.

**Corollary 1** *Assume $G = 2$ and $\dim(Y_{it}) = 1$. Let $\hat{\boldsymbol{\gamma}}_{NR}$ be the estimated group assignments based on sample $\mathcal{R}$, and let $\tilde{\boldsymbol{\mu}}_{NP}(\hat{\boldsymbol{\gamma}}_{NR})$ be the estimated group means from sample $\mathcal{P}$ using group assignments $\hat{\gamma}_{NR}$. Define the t-statistic for the differences in the estimated means as*

$$tstat_{NPR} = \frac{\sqrt{NP}\left(\tilde{\mu}_{1,NP}(\hat{\boldsymbol{\gamma}}_{NR}) - \tilde{\mu}_{2,NP}(\hat{\boldsymbol{\gamma}}_{NR})\right)}{\hat{\omega}_{NPR}} \quad (2)$$

$$\text{where} \quad \hat{\omega}^2_{NPR} \equiv \frac{1}{NP}\sum_{i=1}^{N}\sum_{t\in\mathcal{P}}\left(Y_{it} - \bar{Y}_{iP}\right)^2\left(\hat{\pi}_{1,NR}^{-2}\mathbf{1}\{\hat{\gamma}_{i,NR} = 1\} + \hat{\pi}_{2,NR}^{-2}\mathbf{1}\{\hat{\gamma}_{i,NR} = 2\}\right) \quad (3)$$

$$\bar{Y}_{iP} \equiv \frac{1}{P}\sum_{t\in\mathcal{P}} Y_{it} \quad (4)$$

$$\hat{\pi}_{g,NR} \equiv \frac{1}{N}\sum_{i=1}^{N}\mathbf{1}\{\hat{\gamma}_{i,NR} = g\}, \quad \text{for } g = 1,2 \quad (5)$$

*(a) Under Assumptions 1 and 2,*

$$tstat_{NPR} \xrightarrow{d} N(0,1), \quad \text{as } N, P, R \to \infty \quad (6)$$

*(b) Under Assumptions 1 and 2',*

$$|tstat_{NPR}| \xrightarrow{p} \infty, \quad \text{as } N, P, R \to \infty \quad (7)$$

First, we consider allowing for general time series dependence up to some lag $M$. To do so, we define $\mathcal{G}_t$ as the information set $\sigma\left(\{Y_{is}\}_{i=1}^{N}, s \leq t\right)$, and modify Assumption 1 to:

**Assumption 1″:** (a) The data come from $Y_{it} = m_i + \varepsilon_{it}$, for $i = 1, ..., N$, and $t = 1, .., T$, where $m_i \in [\underline{m}, \bar{m}] \subset \mathbb{R}$ and $V[\varepsilon_{it}] \equiv \sigma_i^2 \in [\underline{\sigma}^2, \bar{\sigma}^2] \subset \mathbb{R}_+ \ \forall \ i$, $E[\varepsilon_{it}] = 0$ and $E\left[|\varepsilon_{it}|^{4+\delta}\right] < \infty$ $\forall \ i$ for some $\delta > 0$, (b) $\varepsilon_{it} \perp\!\!\!\perp \varepsilon_{js} \ \forall \ t, s$, for $i \neq j$ (c) $\varepsilon_{it} \perp\!\!\!\perp X$ for all $X \in \mathcal{G}_{t-M}$, for $\forall \ i, t$ and (d) $N, P, R \to \infty$.

Assumption 1″(a) allows for cross-sectional heteroskedasticity, and heterogeneity more generally, in the distribution of residuals, subject to them being mean zero and having finite fourth moments. Assumption 1″(b) imposes cross-sectional independence, and 1″(c) allows for general time series dependence up to lag $M$, but imposes independence beyond $M$ lags. The main change required when allowing for time series dependence is that the formation of subsamples now requires some structure. We suggest using $\mathcal{R} = \{1, 2, ..., R - M\}$ and $\mathcal{P} = \{R+1, ..., R + P \equiv T\}$.

**Theorem 6** *Let $\hat{\boldsymbol{\gamma}}_{NR}$ be the estimated group assignments based on sample $\mathcal{R}$, and let $\tilde{\boldsymbol{\mu}}_{NP}(\hat{\boldsymbol{\gamma}}_{NR})$ be the estimated group means from sample $\mathcal{P}$ using group assignments $\hat{\boldsymbol{\gamma}}_{NR}$. Define the t-statistic for the differences in the estimated means as*

$$tstat_{NPR} = \frac{\sqrt{NP}\left(\tilde{\mu}_{1,NP}(\hat{\boldsymbol{\gamma}}_{NR}) - \tilde{\mu}_{2,NP}(\hat{\boldsymbol{\gamma}}_{NR})\right)}{\hat{\omega}_{NPR}}$$

$$\text{where } \hat{\omega}_{NPR}^2 \equiv \sum_{i=1}^{N} \boldsymbol{\iota}_P' \hat{\boldsymbol{\varepsilon}}_i \hat{\boldsymbol{\varepsilon}}_i' \boldsymbol{\iota}_P \left(\hat{\pi}_{1,NR}^{-2} \mathbf{1}\{\hat{\gamma}_{i,NR} = 1\} - \hat{\pi}_{2,NR}^{-2} \mathbf{1}\{\hat{\gamma}_{i,NR} = 2\}\right)$$

$$\hat{\boldsymbol{\varepsilon}}_{iP} = \mathbf{Y}_{iP} - \boldsymbol{\iota}_P \bar{Y}_{iP}$$

$$\text{and } \hat{\pi}_{g,NR} \equiv \frac{1}{N}\sum_{i=1}^{N} \mathbf{1}\{\hat{\gamma}_{i,NR} = g\}, \text{ for } g = 1, 2$$

*and $\mathbf{Y}_{iP} \equiv [Y_{i1}, ..., Y_{iP}]'$ and $\boldsymbol{\iota}_P$ is a $P \times 1$ vector of ones.*

*(a) Under Assumptions 1″ and 2,*

$$tstat_{NPR} \xrightarrow{d} N(0,1), \text{ as } N, P \to \infty$$

*(b) Under Assumptions 1″ and 2′,*

$$|tstat_{NPR}| \xrightarrow{p} \infty, \text{ as } N, P, R \to \infty$$

**Proof of Theorem 6.** (a) We first find the limiting distribution of $\sqrt{NP}\left(\tilde{\mu}_{1,NP}(\hat{\gamma}_{NR}) - \tilde{\mu}_{2,NP}(\hat{\gamma}_{NR})\right)$ conditional on $\mathcal{F}_R$. Note that

$$\tilde{\mu}_{g,NP}(\hat{\gamma}_{NR}) = \frac{1}{\hat{N}_{g,NR}} \sum_{i=1}^{N} \left( \mathbf{1}\{\hat{\gamma}_{i,NR} = g\} \frac{1}{P} \sum_{t=R+1}^{T} Y_{i,t} \right)$$

$$\equiv \frac{1}{NP} \sum_{i=1}^{N} \sum_{t=R+1}^{T} Y_{i,t} \hat{\pi}_{g,NR}^{-1} \mathbf{1}\{\hat{\gamma}_{i,NR} = g\}$$

Then

$$\sqrt{NP}\left(\tilde{\mu}_{1,NP}(\hat{\gamma}_{NR}) - \tilde{\mu}_{2,NP}(\hat{\gamma}_{NR})\right) = \frac{1}{\sqrt{NP}} \sum_{i=1}^{N} \sum_{t=R+1}^{T} \hat{Z}_{i,NR} \varepsilon_{it}$$

where $\hat{Z}_{i,NR} \equiv \hat{\pi}_{1,NR}^{-1} \mathbf{1}\{\hat{\gamma}_{i,NR} = 1\} - \hat{\pi}_{2,NR}^{-1} \mathbf{1}\{\hat{\gamma}_{i,NR} = 2\}$. We now verify that we can invoke a CLT for $\sqrt{NP} \frac{1}{NP} \sum_{i=1}^{N} \sum_{t=R+1}^{P} \xi_{it,NR}$, where $\xi_{it,NR} \equiv \hat{Z}_{i,NR} \varepsilon_{it}$. Note that conditional on $\mathcal{F}_R$, the sequence $\{\xi_{it,NR}\}$ is heterogeneously distributed, and $M$-dependent by Assumption 1''(c) which immediately implies strong mixing. Also note that conditional on $\mathcal{F}_R$, $\xi_{it,NR}$ is independent of $\xi_{jt,NR}$ $\forall$ $i \neq j$. Then note that

$$E\left[\xi_{it,NR} | \mathcal{F}_R\right] = \hat{Z}_{i,NR} E\left[\varepsilon_{it} | \mathcal{F}_R\right] = \hat{Z}_{i,NR} E\left[E\left[\varepsilon_{it} | \mathcal{G}_{t-M}\right] | \mathcal{F}_R\right] = 0, \text{ for } t \geq R+1$$

Next, let

$$\boldsymbol{\xi}'_{i,NPR} \equiv \left[\xi_{i,R+1,NR}, ..., \xi_{i,R+P,NR}\right] = \hat{Z}_{i,NR}\left[\varepsilon_{i,R+1}, ..., \varepsilon_{i,R+P}\right]' \equiv \hat{Z}_{i,NR} \boldsymbol{\varepsilon}'_{iP}$$

and note that

$$E\left[\boldsymbol{\xi}_{i,NPR} \boldsymbol{\xi}'_{i,NPR} \middle| \mathcal{F}_R\right] = \hat{Z}_{i,NR}^2 E\left[\boldsymbol{\varepsilon}_{iP} \boldsymbol{\varepsilon}'_{iP}\right] \equiv \hat{Z}_{i,NR}^2 \Omega_i$$

Note that by Assumption 1''(a) and (c), $\Omega_i$ is a Toeplitz matrix, with $\sigma_i^2$ on the main diagonal, $\psi_{i,1} \equiv Cov\left[\varepsilon_{i,t}, \varepsilon_{i,t+1}\right]$ on the secondary diagonal, etc. out to $\psi_{i,M} \equiv Cov\left[\varepsilon_{i,t}, \varepsilon_{i,t+M}\right]$ on the $(M+1)^{th}$ diagonal, and with zeros elsewhere. This structure simplifies the estimation of $\Omega_i$.

Finally, define

$$\bar{\omega}_{NR}^2 \equiv \lim_{N,P \to \infty} \frac{1}{NP} \sum_{i=1}^{N} \hat{Z}_{i,NR}^2 \boldsymbol{\iota}'_P \Omega_i \boldsymbol{\iota}_P$$

$$= \lim_{N,P \to \infty} \frac{1}{N} \sum_{i=1}^{N} \hat{Z}_{i,NR}^2 \sigma_i^2 + \lim_{N,P \to \infty} \frac{2}{N} \sum_{i=1}^{N} \hat{Z}_{i,NR}^2 \left( \sum_{k=1}^{M} (1 - k/P) \psi_{i,k} \right)$$

The general estimator of the asymptotic covariance in Hansen (2007) is given below, which we then simplify based on our $M$-dependence assumption.

$$\hat{\omega}_{NPR}^2 = \frac{1}{N}\sum_{i=1}^{N}\hat{Z}_{i,NR}^2\hat{\psi}_{i,0,P} + \frac{2}{N}\sum_{i=1}^{N}\hat{Z}_{i,NR}^2\left(\sum_{k=1}^{M}(1-k/P)\hat{\psi}_{i,k,P}\right)$$

$$\text{where } \hat{\psi}_{i,k,P} \equiv \frac{1}{P}\sum_{t=R+1}^{T-k}\left(Y_{it}-\bar{Y}_{iP}\right)\left(Y_{i,t+k}-\bar{Y}_{iP}\right), \quad k=0,1,...,M$$

Our Assumption 1″ is sufficient for Assumptions 1, 2 and 3(b) of Hansen (2007, Theorem 3), and thus we have, conditional on $\mathcal{F}_R$,

$$\frac{\sqrt{NP}\frac{1}{NP}\sum_{i=1}^{N}\sum_{t=R+1}^{T}\hat{Z}_{i,NR}\varepsilon_{it}}{\bar{\omega}_{NR}} \xrightarrow{d} N(0,1) \text{ and } \hat{\omega}_{NPR}^2 \xrightarrow{p} \bar{\omega}_{NR}^2 \text{ as } N,P,R\to\infty$$

This implies that the $t$-statistic obeys

$$\frac{\sqrt{NP}\frac{1}{NP}\sum_{i=1}^{N}\sum_{t=R+1}^{T}\hat{Z}_{i,NR}\varepsilon_{it}}{\hat{\omega}_{NPR}} \xrightarrow{d} N(0,1), \text{ as } N,P,R\to\infty$$

As the limiting distribution of the $t$-statistic does not depend on $\mathcal{F}_R$, its unconditional distribution is also $N(0,1)$, completing the proof.

(b) As in Theorem 1(b), note that $\tilde{\mu}_{NP}(\hat{\gamma}_{NR}) - \mu^* = (\hat{\mu}_{NR} - \mu^*) + (\tilde{\mu}_{NP}(\hat{\gamma}_{NR}) - \hat{\mu}_{NR})$. Our Assumption 1″ is sufficient for Assumption 1 of Bonhomme and Manresa (2015), and their Theorem 1 implies that the first term on the RHS is $o_p(1)$, as $N,R\to\infty$. The second term is:

$$\begin{aligned}
\tilde{\mu}_{g,NP}(\hat{\gamma}_{NR}) - \hat{\mu}_{g,NR} &= \frac{1}{N}\sum_{i=1}^{N}\hat{\pi}_{g,NR}^{-1}\mathbf{1}\{\hat{\gamma}_{i,NR}=g\}\left(\frac{1}{P}\sum_{t=R+1}^{T}Y_{i,t} - \frac{1}{R-M}\sum_{t=1}^{R-M}Y_{i,t}\right) \\
&= \frac{1}{N}\sum_{i=1}^{N}\hat{\pi}_{g,NR}^{-1}\mathbf{1}\{\hat{\gamma}_{i,NR}=g\}\left(\frac{1}{P}\sum_{t=R+1}^{T}\varepsilon_{i,t} - \frac{1}{R-M}\sum_{t=1}^{R-M}\varepsilon_{i,t}\right) \\
&\leq \underline{\pi}^{-1}\frac{1}{N}\sum_{i=1}^{N}\left(\frac{1}{P}\sum_{t=R+1}^{T}\varepsilon_{i,t} - \frac{1}{R-M}\sum_{t=1}^{R-M}\varepsilon_{i,t}\right) \\
&= \underline{\pi}^{-1}\left(\frac{1}{NP}\sum_{i=1}^{N}\sum_{t=R+1}^{T}\varepsilon_{i,t} - \frac{1}{N(R-M)}\sum_{i=1}^{N}\sum_{t=1}^{R-M}\varepsilon_{i,t}\right) \\
&= o_p(1), \text{ as } N,P,R\to\infty
\end{aligned}$$

since our Assumption 1″ is sufficient for Assumptions 1, 2 and 3(b) of Hansen (2007), which implies that $\frac{1}{N(R-M)}\sum_{i=1}^{N}\sum_{t=1}^{R-M}\varepsilon_{i,t} = o_p(1)$ and $\frac{1}{NP}\sum_{i=1}^{N}\sum_{t=R+1}^{T}\varepsilon_{i,t} = o_p(1)$. This holds for $g=1,2$, and thus $\tilde{\mu}_{NP}(\hat{\gamma}_{NR}) \xrightarrow{p} \mu^*$, as $N, P, R \to \infty$. This implies that $\tilde{\mu}_{1,NP}(\hat{\gamma}_{NR}) - \tilde{\mu}_{2,NP}(\hat{\gamma}_{NR}) \xrightarrow{p} \mu_1^* - \mu_2^* \neq 0$ by Assumption 2'(b). Thus $|tstat| \xrightarrow{p} \infty$, as $N, P, R \to \infty$. ∎

We next consider allowing for general time series and cross-sectional dependence, by adapting Assumption 1 of Bonhomme and Manresa (2015) to our application. Consider the following assumption. Let $K$ denote some finite constant.

**Assumption 1'''**: (a) The data come from $Y_{it} = m_i + \varepsilon_{it}$, for $i = 1, ..., N$, and $t = 1, .., T$, where $m_i \in [\underline{m}, \bar{m}] \subset \mathbb{R}$ and $V[\varepsilon_{it}] \equiv \sigma_i^2 \in [\underline{\sigma}^2, \bar{\sigma}^2] \subset \mathbb{R}_+$ $\forall i$, and $E[\eta_{it}^4] \leq \bar{\kappa} < \infty$ $\forall i$

(b) $\left| \frac{1}{NT} \sum_{i=1}^{N} \sum_{t=1}^{T} \sum_{s=1}^{T} E[\varepsilon_{it}\varepsilon_{is}] \right| \leq K < \infty$

(c) $\left| \frac{1}{N^2 T} \sum_{i=1}^{N} \sum_{j=1}^{N} \sum_{t=1}^{T} \sum_{s=1}^{T} Cov[\varepsilon_{it}\varepsilon_{jt}, \varepsilon_{is}\varepsilon_{js}] \right| \leq K < \infty$

(d) $\frac{1}{N} \sum_{i=1}^{N} \sum_{j=1}^{N} \left| \frac{1}{T} \sum_{t=1}^{T} E[\varepsilon_{it}\varepsilon_{jt}] \right| \leq K < \infty$

(e) $\sqrt{NT} \frac{1}{NT} \sum_{i=1}^{N} \sum_{t=1}^{T} \varepsilon_{it} \xrightarrow{d} N(0, \bar{v}^2)$ for some $\bar{v}^2 > 0$, and there exists an estimator $\hat{v}_{NT}^2$ that is robust to cross-sectional heteroskedasticity in $\{\varepsilon_{it}\}$ and is consistent for $\bar{v}^2$, as $N, T \to \infty$.

(f) $N, P, R \to \infty$.

Assumption 1'''(a) allows for cross-sectional heteroskedasticity, and heterogeneity more generally, in the distribution of residuals, subject to them being mean zero and having finite fourth moments. Assumptions 1'''(b) and (c) imposes restrictions on the amount of time series dependence in the data, and 1'''(d) limits the amount of cross-sectional dependence. Assumption 1'''(e) is a high level assumption that a CLT can be invoked for the sample average of $\{\varepsilon_{it}\}$, and that a consistent estimator of the asymptotic variance is available. There are a variety of CLTs and LLNs that can be used in panel applications to satisfy this assumption, see Pesaran (2015) for a recent textbook treatment of this area. The requirement that this estimator is robust to cross-sectional heteroskedasticity is a mild requirement and is satisfied by many estimators in the literature.

**Theorem 7** *Let $\hat{\gamma}_{NR}$ be the estimated group assignments based on sample $\mathcal{R}$, and let $\tilde{\mu}_{NP}(\hat{\gamma}_{NR})$ be the estimated group means from sample $\mathcal{P}$ using group assignments $\hat{\gamma}_{NR}$. Define the t-statistic* for the differences in the estimated means as

$$tstat_{NPR} = \frac{\sqrt{NP}\left(\tilde{\mu}_{1,NP}\left(\hat{\gamma}_{NR}\right) - \tilde{\mu}_{2,NP}\left(\hat{\gamma}_{NR}\right)\right)}{\hat{\omega}_{NPR}}$$

where $\hat{\omega}_{NPR}^2$ is an estimator of the asymptotic variance of

$$\xi_{it,NR} \equiv \left(\hat{\pi}_{1,NR}^{-1}\mathbf{1}\left\{\hat{\gamma}_{i,NR} = 1\right\} - \hat{\pi}_{2,NR}^{-1}\mathbf{1}\left\{\hat{\gamma}_{i,NR} = 2\right\}\right)\varepsilon_{it}$$

and takes the same functional form as the estimator $\hat{v}_{NT}^2$ in Assumption 1'''(e).

(a) Under Assumptions 1''' and 2,

$$tstat_{NPR} \xrightarrow{d} N(0,1), \quad as\ N, P \to \infty \tag{8}$$

(b) Under Assumptions 1''' and 2',

$$|tstat_{NPR}| \xrightarrow{p} \infty, \quad as\ N, P, R \to \infty \tag{9}$$

**Proof of Theorem 7.** (a) Note that

$$\sqrt{NP}\left(\tilde{\mu}_{1,NP}\left(\hat{\gamma}_{NR}\right) - \tilde{\mu}_{2,NP}\left(\hat{\gamma}_{NR}\right)\right) = \frac{1}{\sqrt{NP}}\sum_{i=1}^{N}\sum_{t\in\mathcal{P}}\varepsilon_{it}\left(\hat{\pi}_{1,NR}^{-1}\mathbf{1}\left\{\hat{\gamma}_{i,NR} = 1\right\} - \hat{\pi}_{2,NR}^{-1}\mathbf{1}\left\{\hat{\gamma}_{i,NR} = 2\right\}\right)$$

$$= \frac{1}{\sqrt{NP}}\sum_{i=1}^{N}\sum_{t\in\mathcal{P}}\hat{Z}_{i,NR}\varepsilon_{it}$$

where $\hat{Z}_{i,NR} \equiv \hat{\pi}_{1,NR}^{-1}\mathbf{1}\left\{\hat{\gamma}_{i,NR} = 1\right\} - \hat{\pi}_{2,NR}^{-1}\mathbf{1}\left\{\hat{\gamma}_{i,NR} = 2\right\}$. By Assumption 1'''(e) we know that $\frac{1}{\sqrt{NP}}\sum_{i=1}^{N}\sum_{t=1}^{P}\varepsilon_{it} \xrightarrow{d} N(0,\bar{v}^2)$, so

$$\bar{v}^2 = \lim_{N,P\to\infty}V\left[\frac{1}{\sqrt{NP}}\sum_{i=1}^{N}\sum_{t=1}^{P}\varepsilon_{it}\right] = \lim_{N,P\to\infty}\frac{1}{NP}\sum_{i=1}^{N}\sum_{j=1}^{N}\sum_{t=1}^{P}\sum_{s=1}^{P}E\left[\varepsilon_{it}\varepsilon_{js}\right]$$

Conditional on $\mathcal{F}_R$, the weights, $\hat{Z}_{i,NR}$, on $\varepsilon_{it}$ are known, and are bounded since $\underline{\pi} > 0$. Define the variable $\xi_{it,NR} \equiv \hat{Z}_{i,NR}\varepsilon_{it}$, and note that we have:

$$E\left[\xi_{it,NR}\big|\mathcal{F}_R\right] = \hat{Z}_{i,NR}E\left[\varepsilon_{it}\right] = 0$$

Moreover,

$$E\left[\left|\xi_{it,NR}\right|^q \big| \mathcal{F}_R\right] = \left|\hat{Z}_{i,NR}\right|^q E\left[|\varepsilon_{it}|^q\right], \text{ for } q \text{ s.t. } E\left[|\varepsilon_{it}|^q\right] < \infty$$

$$E\left[\xi_{it,NR}\xi_{is,NR} \big| \mathcal{F}_R\right] = \hat{Z}_{i,NR}^2 E\left[\varepsilon_{it}\varepsilon_{is}\right] \ \forall\ i,t,s$$

$$E\left[\xi_{it,NR}\xi_{jt,NR} \big| \mathcal{F}_R\right] = \hat{Z}_{i,NR}\hat{Z}_{j,NR} E\left[\varepsilon_{it}\varepsilon_{jt}\right] \ \forall\ i,j,t$$

$$Cov\left[\xi_{it,NR}\xi_{jt,NR},\xi_{jt,NR}\xi_{js,NR} \big| \mathcal{F}_R\right] = \hat{Z}_{i,NR}^2 \hat{Z}_{j,NR}^2 Cov\left[\varepsilon_{it}\varepsilon_{jt},\varepsilon_{is}\varepsilon_{js}\right] \ \forall\ i,j,t,s$$

and so the moment and memory properties of $\{\xi_{it,NR}\}$ are completely determined by the moment and memory properties of $\{\varepsilon_{it}\}$. Thus any CLT that applies to $\{\varepsilon_{it}\}$, and which allows for cross-sectional heteroskedasticity, will also apply to $\{\xi_{it,NR}\}$, conditional on $\mathcal{F}_R$. This implies that

$$\frac{1}{\sqrt{NP}}\sum_{i=1}^{N}\sum_{t=1}^{P}\xi_{it,NR} \xrightarrow{d} N\left(0,\bar{\omega}^2\right)$$

$$\text{where} \quad \bar{\omega}^2 = \lim_{N,P\to\infty} V\left[\frac{1}{\sqrt{NP}}\sum_{i=1}^{N}\sum_{t=1}^{P}\xi_{it,NR}\right]$$

By Assumption $1'''$(e) we know that there exists an estimator $\hat{v}_{NP}^2$ such that $\hat{v}_{NP}^2 \xrightarrow{p} \bar{v}^2$, as $N, P \to \infty$. As $\hat{Z}_{i,NR}$ is non-zero and finite, any estimator $\hat{v}_{NP}^2$ that is consistent for $\bar{v}^2$, and robust to cross-sectional heteroskedasticity, can also be applied to $\xi_{it,NR}$, yielding an estimator $\hat{\omega}_{NPR}^2$ that is consistent for $\bar{\omega}^2$. This implies that the $t$-statistic obeys:

$$tstat = \frac{\sqrt{NP}\left(\tilde{\mu}_{1,NP}\left(\hat{\gamma}_{NR}\right) - \tilde{\mu}_{2,NP}\left(\hat{\gamma}_{NR}\right)\right)}{\hat{\omega}_{NPR}} \xrightarrow{d} N\left(0,1\right) \ \text{ as } N,P,R \to \infty$$

As the limiting distribution of the $t$-statistic does not depend on $\mathcal{F}_R$, its unconditional distribution is also $N\left(0,1\right)$, completing the proof.

(b) Note that $\tilde{\mu}_{NP}\left(\hat{\gamma}_{NR}\right) - \mu^* = \left(\hat{\mu}_{NR} - \mu^*\right) + \left(\tilde{\mu}_{NP}\left(\hat{\gamma}_{NR}\right) - \hat{\mu}_{NR}\right)$. Our Assumption $1'''$ is sufficient for Assumption 1 of Bonhomme and Manresa (2015), and their Theorem 1 implies that the first term on the RHS is $o_p(1)$, as $N, R \to \infty$. The second term is:

$$\begin{aligned}
\tilde{\mu}_{g,NP}(\hat{\gamma}_{NR}) - \hat{\mu}_{g,NR} &= \frac{1}{N}\sum_{i=1}^{N}\hat{\pi}_{g,NR}^{-1}\mathbf{1}\{\hat{\gamma}_{i,NR} = g\}\left(\frac{1}{P}\sum_{t\in\mathcal{P}}Y_{i,t} - \frac{1}{R}\sum_{t\in\mathcal{R}}Y_{i,t}\right) \\
&= \frac{1}{N}\sum_{i=1}^{N}\hat{\pi}_{g,NR}^{-1}\mathbf{1}\{\hat{\gamma}_{i,NR} = g\}\left(\frac{1}{P}\sum_{t\in\mathcal{P}}\varepsilon_{i,t} - \frac{1}{R}\sum_{t\in\mathcal{R}}\varepsilon_{i,t}\right) \\
&\leq \underline{\pi}^{-1}\left(\frac{1}{NP}\sum_{i=1}^{N}\sum_{t\in\mathcal{P}}\varepsilon_{i,t} - \frac{1}{NR}\sum_{i=1}^{N}\sum_{t\in\mathcal{R}}\varepsilon_{i,t}\right) \\
&= o_p(1), \text{ as } N, P, R \to \infty
\end{aligned}$$

since $\underline{\pi} > 0$ and using a LLN for $\frac{1}{NP}\sum_{i=1}^{N}\sum_{t\in\mathcal{P}}\varepsilon_{i,t}$ and $\frac{1}{NR}\sum_{i=1}^{N}\sum_{t\in\mathcal{R}}\varepsilon_{i,t}$ which follows from Theorem 1 of Bonhomme and Manresa (2015). This holds for $g = 1, 2$, and thus $\tilde{\mu}_{NP}(\hat{\gamma}_{NR}) \xrightarrow{p} \mu^*$, as $N, P, R \to \infty$. This implies that $\tilde{\mu}_{1,NP}(\hat{\gamma}_{NR}) - \tilde{\mu}_{2,NP}(\hat{\gamma}_{NR}) \xrightarrow{p} \mu_1^* - \mu_2^* \neq 0$ by Assumption 2'(b). Thus $|tstat| \xrightarrow{p} \infty$, as $N, P, R \to \infty$. ∎

## S.A.2: Additional proofs

**Proof of Lemma 1.** We know that the limit of the objective function of the correctly specified model is minimized at $(\boldsymbol{\mu}^*, \boldsymbol{\gamma}^*)$, and the MSE at that point is

$$\begin{aligned}
MSE^*(\boldsymbol{\mu}^*, \boldsymbol{\gamma}^*) &= \lim_{N,T\to\infty}\frac{1}{NT}\sum_{i=1}^{N}\sum_{t=1}^{T}\sum_{g=1}^{G}(Y_{it} - \mu_g^*)^2\mathbf{1}\{\gamma_i^* = g\} \\
&= \lim_{N,T\to\infty}\frac{1}{NT}\sum_{i=1}^{N}\sum_{t=1}^{T}\varepsilon_{it}^2 \\
&= \frac{1}{N}\sum_{i=1}^{N}\sigma_i^2 \\
&\equiv \bar{\sigma}^2
\end{aligned}$$

Let $\boldsymbol{\gamma}^{\star}$ be such that, for all $i, j \in \{1, ..., N\}$, $\gamma_i^{\star} = \gamma_j^{\star} \Rightarrow \gamma_i^* = \gamma_j^*$. That is, all clusters defined by $\boldsymbol{\gamma}^{\star}$ can be generated by taking the correct set of clusters (given by $\boldsymbol{\gamma}^*$) and then splitting some $\boldsymbol{\gamma}^*$-clusters into two or more clusters. This implies that $\boldsymbol{\mu}_g^{\star} = \boldsymbol{\mu}_{g'}^*$ for some $g'$, for all $g$. For any such $(\boldsymbol{\mu}^{\star}, \boldsymbol{\gamma}^{\star})$ the limit of the objective function is

$$MSE^{*}\left(\boldsymbol{\mu}^{\star}, \boldsymbol{\gamma}^{\star}\right) = \lim_{N,T\to\infty} \frac{1}{NT} \sum_{i=1}^{N} \sum_{t=1}^{T} \sum_{g=1}^{\tilde{G}} \left(Y_{it} - \mu_{g}^{\star}\right)^{2} \mathbf{1}\left\{\gamma_{i}^{\star} = g\right\} = \lim_{N,T\to\infty} \frac{1}{NT} \sum_{i=1}^{N} \sum_{t=1}^{T} \varepsilon_{it}^{2}$$

$$= \frac{1}{N} \sum_{i=1}^{N} \sigma_{i}^{2} \equiv \bar{\sigma}^{2} \leq MSE^{*}\left(\boldsymbol{\mu}, \boldsymbol{\gamma}\right)$$

and so $(\boldsymbol{\mu}^{\star}, \boldsymbol{\gamma}^{\star})$ is a solution to the population $\tilde{G}$-cluster estimation problem. ■

**Lemma 3** For $d = 1$, Assumption 2'(b) implies Assumption 3''(b).

**Proof of Lemma 3.** Consider the case that $G = 3$ and $\tilde{G} = 2$ for simplicity, and assume $\mu_{1}^{*} < \mu_{2}^{*} < \mu_{3}^{*}$. Every element of a group has the same mean (by Assumption 2') and so if it is optimal for one member of a given group to be assigned to a specific group in the $\tilde{G}$-cluster model then it is optimal for *all* members of that true group. This implies that there are no split true groups between the $\tilde{G}$-cluster model groups. There are then three possible groupings for the $\tilde{G} = 2$ model, in terms of the true group assignments: $\{1, (2,3)\}$, $\{(1,2), 3\}$, $\{(1,3), 2\}$. The latter allocation can be easily shown to be suboptimal since $\mu_{1}^{*} < \mu_{2}^{*} < \mu_{3}^{*}$, so we need only consider the first two cases.

In the first case, we have $\mu_{1}^{\star} = \mu_{1}^{*}$, since that group comprises all the true group one variables. The other location parameter will be a convex combination of $\mu_{2}^{*}$ and $\mu_{3}^{*}$:

$$\mu_{2}^{\star} = \frac{\pi_{2}}{\pi_{2} + \pi_{3}} \mu_{2}^{*} + \frac{\pi_{3}}{\pi_{2} + \pi_{3}} \mu_{3}^{*}$$

Then note that $\left|\mu_{1}^{\star} - \mu_{2}^{\star}\right| = \left|\mu_{1}^{*} - \mu_{2}^{\star}\right| > |\mu_{1}^{*} - \mu_{2}^{*}| > c$, where the first inequality holds since $\mu_{2}^{\star} \in (\mu_{2}^{*}, \mu_{3}^{*})$ and the second inequality holds by Assumption 2'(b). A similar inequality holds if we consider the other allocation: $\left|\mu_{1}^{\star} - \mu_{2}^{\star}\right| = \left|\mu_{1}^{\star} - \mu_{3}^{*}\right| > |\mu_{2}^{*} - \mu_{3}^{*}| > c$ since in this case we have $\mu_{2}^{\star} = \mu_{3}^{*}$ and $\mu_{1}^{\star} \in (\mu_{1}^{*}, \mu_{2}^{*})$. The extension to the general case $G > \tilde{G} \geq 2$ is proven similarly.

Next we provide an example where this implication fails for $d > 1$. Consider $d = 2$, $G = 3$ and $\tilde{G} = 2$. Assume $\boldsymbol{\mu}_{1}^{*} = [0,0]$, $\boldsymbol{\mu}_{2}^{*} = [2,0]$ and $\boldsymbol{\mu}_{3}^{*} = \left[1, \sqrt{3}\right]$, i.e., these points form an equilateral triangle on $\mathbb{R}^2$ with side lengths equal to two. Assume that $\pi_1 = \pi_2 \geq \underline{\pi} > 0$ and $\pi_3 > 1/3$, leading to the optimal $\tilde{G} = 2$ group assignment being $\{(1,2), 3\}$. Thus $\boldsymbol{\mu}_2^\star = \boldsymbol{\mu}_3^*$ and

$$\boldsymbol{\mu}_1^\star = \frac{1}{2}(\boldsymbol{\mu}_1^* + \boldsymbol{\mu}_2^*) = [1, 0]$$

In this case we find $\left\|\boldsymbol{\mu}_1^\star - \boldsymbol{\mu}_2^\star\right\| = \sqrt{3} < \min_{g \neq g'} \left\|\boldsymbol{\mu}_g^* - \boldsymbol{\mu}_{g'}^*\right\|$. Thus the $\tilde{G} = 2$ model has optimal clusters that are closer together than the clusters in the DGP. ∎

**Proof of Theorem 2.** (a) This case is identical to the case considered in Theorem 1(a): a model with $\tilde{G}$ clusters is estimated, but the null of only a single cluster is true. Thus we obtain

$$Fstat \xrightarrow{d} \chi^2_{\tilde{G}-1}, \text{ as } N, P, R \to \infty.$$

(b) Now we consider a $\tilde{G}$-cluster model when the DGP has $G \in \{2, ..., \tilde{G}-1\}$ clusters, and so the $\tilde{G}$-cluster model is too large. Note that

$$\tilde{\boldsymbol{\mu}}_{NP}(\hat{\boldsymbol{\gamma}}_{NR}) - \boldsymbol{\mu}^\star = \left(\hat{\boldsymbol{\mu}}_{NR} - \boldsymbol{\mu}^\star\right) + \left(\tilde{\boldsymbol{\mu}}_{NP}(\hat{\boldsymbol{\gamma}}_{NR}) - \hat{\boldsymbol{\mu}}_{NR}\right)$$

The first term on the RHS is $o_p(1)$ as $N, R \to \infty$ by Assumption 3'(a). The second term is treated as in Theorem 1(b) and is $o_p(1)$ as $N, P, R \to \infty$.

By Lemma 1, $\boldsymbol{\mu}^\star$ is a re-ordering of $[\boldsymbol{\mu}^{*\prime}, \boldsymbol{\varphi}^{*\prime}]'$, where $\boldsymbol{\varphi}^*$ is a $\left(\tilde{G} - G\right)$ vector with elements drawn with replacement from $\boldsymbol{\mu}^*$. The well-separatedness assumption on the DGP (Assumption 2'(b)) implies that all of the $G(G-1)/2$ pairwise differences of elements of $\boldsymbol{\mu}^*$ are non-zero, i.e., $\left|\mu_g^* - \mu_{g'}^*\right| > c > 0 \ \forall \ g \neq g'$. Combining this with Lemma 1 we have:

$$\sum_{g=1}^{\tilde{G}-1} \sum_{g'=g+1}^{\tilde{G}} \mathbf{1}\left\{\left|\mu_g^\star - \mu_{g'}^\star\right| = 0\right\} \leq \left(\tilde{G} - G + 1\right)\left(\tilde{G} - G\right)/2$$

and so $$\sum_{g=1}^{\tilde{G}-1} \sum_{g'=g+1}^{\tilde{G}} \mathbf{1}\left\{\left|\mu_g^\star - \mu_{g'}^\star\right| > c\right\} \geq \left(4\tilde{G} - 3G\right)(G-1)/2$$

Thus, while not all of the pairwise differences in $\boldsymbol{\mu}_g^\star$ will be non-zero, there will be at least $\left(4\tilde{G} - 3G\right)(G-1)/2$ non-zero pairwise differences. This implies that

$$\tilde{\boldsymbol{\mu}}'_{NP}(\hat{\boldsymbol{\gamma}}_{NR}) A'_{1,\tilde{G}} \left(A_{1,\tilde{G}} \hat{\Omega}_{NPR} A'_{1,\tilde{G}}\right)^{-1} A_{1,\tilde{G}} \tilde{\boldsymbol{\mu}}_{NP}(\hat{\boldsymbol{\gamma}}_{NR}) \xrightarrow{p} \boldsymbol{\mu}^{\star\prime} A'_{1,\tilde{G}} \left(A_{1,\tilde{G}} \bar{\Omega}_{NR} A'_{1,\tilde{G}}\right)^{-1} A_{\tilde{G}} \boldsymbol{\mu}^\star > 0$$

by the positive definiteness of $\bar{\Omega}_{NR}$ and the full row rank of $A_G$. And thus

$$Fstat = NP\tilde{\boldsymbol{\mu}}'_{NP}(\hat{\boldsymbol{\gamma}}_{NR}) A'_{1,\tilde{G}} \left( A_{1,\tilde{G}} \hat{\Omega}_{NPR} A'_{1,\tilde{G}} \right)^{-1} A_{1,\tilde{G}} \tilde{\boldsymbol{\mu}}_{NP}(\hat{\boldsymbol{\gamma}}_{NR}) \xrightarrow{p} \infty, \text{ as } N, P, R \to \infty$$

completing the proof.

(c) Now we consider a $\tilde{G}$-cluster model when the DGP has $G > \tilde{G}$ clusters, and so the $\tilde{G}$-cluster model is misspecified. Note that

$$\tilde{\boldsymbol{\mu}}_{NP}(\hat{\boldsymbol{\gamma}}_{NR}) - \boldsymbol{\mu}^{\star} = \left( \hat{\boldsymbol{\mu}}_{NR} - \boldsymbol{\mu}^{\star} \right) + \left( \tilde{\boldsymbol{\mu}}_{NP}(\hat{\boldsymbol{\gamma}}_{NR}) - \hat{\boldsymbol{\mu}}_{NR} \right)$$

The first term on the RHS is $o_p(1)$ as $N, R \to \infty$ by Assumption 3''(a). The second term is treated as in Theorem 1(b) and is $o_p(1)$ as $N, P, R \to \infty$. This implies that

$$\tilde{\boldsymbol{\mu}}'_{NP}(\hat{\boldsymbol{\gamma}}_{NR}) A'_{1,\tilde{G}} \left( A_{1,\tilde{G}} \hat{\Omega}_{NPR} A'_{1,\tilde{G}} \right)^{-1} A_{1,\tilde{G}} \tilde{\boldsymbol{\mu}}_{NP}(\hat{\boldsymbol{\gamma}}_{NR}) \xrightarrow{p} \boldsymbol{\mu}^{\star\prime} A'_{1,\tilde{G}} \left( A_{1,\tilde{G}} \Omega_{NR} A'_{1,\tilde{G}} \right)^{-1} A_{1,\tilde{G}} \boldsymbol{\mu}^{\star} > 0$$

by Assumption 3''(b), the positive definiteness of $\bar{\Omega}_{NP}$ and the full row rank of $A_G$. Thus

$$Fstat = NP\tilde{\boldsymbol{\mu}}'_{NP}(\hat{\boldsymbol{\gamma}}_{NR}) A'_{1,\tilde{G}} \left( A_{1,\tilde{G}} \hat{\Omega}_{NPR} A'_{1,\tilde{G}} \right)^{-1} A_{1,\tilde{G}} \tilde{\boldsymbol{\mu}}_{NP}(\hat{\boldsymbol{\gamma}}_{NR}) \xrightarrow{p} \infty, \text{ as } N, P, R \to \infty$$

completing the proof. ∎

# References

[1] Bonhomme, S., and Manresa, E., 2015, Grouped patterns of heterogeneity in panel data, *Econometrica*, 83(3), 1147-1184.

[2] Hansen, C.B., 2007, Asymptotic properties of a robust variance matrix estimator for panel data when $T$ is large, *Journal of Econometrics*, 141, 597-620.

[3] Pesaran, M.H., 2015, *Time Series and Panel Data Econometrics*, Oxford University Press, Oxford.

Table SA.1: Finite sample rejection rates, no sample splitting

| d | G | N = 30, T = 50 | 30, 250 | 30, 1000 | 150, 50 | 150, 250 | 150, 1000 | 600, 50 | 600, 250 | 600, 1000 |
|---|---|---|---|---|---|---|---|---|---|---|
| | | *Normal data* | | | | | | | | |
| 1 | 2 | 1.000 | 1.000 | 1.000 | 1.000 | 1.000 | 1.000 | 1.000 | 1.000 | 1.000 |
| 1 | 3 | 1.000 | 1.000 | 1.000 | 1.000 | 1.000 | 1.000 | 1.000 | 1.000 | 1.000 |
| 1 | 4 | 1.000 | 1.000 | 1.000 | 1.000 | 1.000 | 1.000 | 1.000 | 1.000 | 1.000 |
| 1 | 5 | 1.000 | **0.999** | 1.000 | 1.000 | 1.000 | 1.000 | 1.000 | 1.000 | 1.000 |
| 1 | Bonf. | 1.000 | 1.000 | 1.000 | 1.000 | 1.000 | 1.000 | 1.000 | 1.000 | 1.000 |
| 2 | 2 | 1.000 | 1.000 | 1.000 | 1.000 | 1.000 | 1.000 | 1.000 | 1.000 | 1.000 |
| 2 | 3 | 1.000 | 1.000 | 1.000 | 1.000 | 1.000 | 1.000 | 1.000 | 1.000 | 1.000 |
| 2 | 4 | 1.000 | 1.000 | 1.000 | 1.000 | 1.000 | 1.000 | 1.000 | 1.000 | 1.000 |
| 2 | 5 | 1.000 | 1.000 | 1.000 | 1.000 | 1.000 | 1.000 | 1.000 | 1.000 | 1.000 |
| 2 | Bonf. | 1.000 | 1.000 | 1.000 | 1.000 | 1.000 | 1.000 | 1.000 | 1.000 | 1.000 |
| 5 | 2 | 1.000 | 1.000 | 1.000 | 1.000 | 1.000 | 1.000 | 1.000 | 1.000 | 1.000 |
| 5 | 3 | 1.000 | 1.000 | 1.000 | 1.000 | 1.000 | 1.000 | 1.000 | 1.000 | 1.000 |
| 5 | 4 | 1.000 | 1.000 | 1.000 | 1.000 | 1.000 | 1.000 | 1.000 | 1.000 | 1.000 |
| 5 | 5 | 1.000 | 1.000 | 1.000 | 1.000 | 1.000 | 1.000 | 1.000 | 1.000 | 1.000 |
| 5 | Bonf. | 1.000 | 1.000 | 1.000 | 1.000 | 1.000 | 1.000 | 1.000 | 1.000 | 1.000 |
| | | *Heterogeneous data* | | | | | | | | |
| 1 | 2 | 1.000 | 1.000 | 1.000 | 1.000 | 1.000 | 1.000 | 1.000 | 1.000 | 1.000 |
| 1 | 3 | 1.000 | 1.000 | 1.000 | 1.000 | 1.000 | 1.000 | 1.000 | 1.000 | 1.000 |
| 1 | 4 | 1.000 | 1.000 | 1.000 | 1.000 | 1.000 | 1.000 | 1.000 | 1.000 | 1.000 |
| 1 | 5 | 1.000 | 1.000 | 1.000 | 1.000 | 1.000 | 1.000 | 1.000 | 1.000 | 1.000 |
| 1 | Bonf. | 1.000 | 1.000 | 1.000 | 1.000 | 1.000 | 1.000 | 1.000 | 1.000 | 1.000 |
| 2 | 2 | 1.000 | 1.000 | 1.000 | 1.000 | 1.000 | 1.000 | 1.000 | 1.000 | 1.000 |
| 2 | 3 | 1.000 | 1.000 | 1.000 | 1.000 | 1.000 | 1.000 | 1.000 | 1.000 | 1.000 |
| 2 | 4 | 1.000 | 1.000 | 1.000 | 1.000 | 1.000 | 1.000 | 1.000 | 1.000 | 1.000 |
| 2 | 5 | 1.000 | 1.000 | 1.000 | 1.000 | 1.000 | 1.000 | 1.000 | 1.000 | 1.000 |
| 2 | Bonf. | 1.000 | 1.000 | 1.000 | 1.000 | 1.000 | 1.000 | 1.000 | 1.000 | 1.000 |
| 5 | 2 | 1.000 | 1.000 | 1.000 | 1.000 | 1.000 | 1.000 | 1.000 | 1.000 | 1.000 |
| 5 | 3 | 1.000 | 1.000 | 1.000 | 1.000 | 1.000 | 1.000 | 1.000 | 1.000 | 1.000 |
| 5 | 4 | 1.000 | 1.000 | 1.000 | 1.000 | 1.000 | 1.000 | 1.000 | 1.000 | 1.000 |
| 5 | 5 | 1.000 | 1.000 | 1.000 | 1.000 | 1.000 | 1.000 | 1.000 | 1.000 | 1.000 |
| 5 | Bonf. | 1.000 | 1.000 | 1.000 | 1.000 | 1.000 | 1.000 | 1.000 | 1.000 | 1.000 |

Notes: This table presents the proportion of simulations in which the null of a single cluster is rejected in favor of multiple clusters, using the test proposed in Theorem 1 but *without* sample splitting, at a 0.05 significance level. The top panel presents results for *iid* Normal data; the lower panel presents results when the distribution is randomly drawn from one of $N(0,1)$, $Exp(2)$, $Unif(-3,3)$, $\chi^2(4)$ or $t(5)$, each standardized to have zero mean and unit variance. The dimension of the variables is denoted $d$, the number of groups considered under the alternative is denoted $G$, the number of variables is denoted $N$, and the number of time series observations is denoted $T$. Rows labeled "Bonf." use tests with a Bonferroni correction to consider $G \in \{2,3,4,5\}$ under the alternative. The number of simulations is 1000.